\documentclass[aps,prb,showpacs,groupedaddress,twocolumn]{revtex4}
\usepackage{graphicx}
\usepackage{verbatim}
\usepackage{mathrsfs}
\usepackage{subfigure}
 \usepackage{gensymb}
\usepackage{color}
\def\be{\begin{equation}}
\def\ee{\end{equation}}
\def\bea{\begin{eqnarray}}
\def\eea{\end{eqnarray}}
\def\bw{\begin{widetext}}
\def\ew{\end{widetext}}
\def\nn{\nonumber}

\usepackage{amsmath,amsfonts,amssymb}

\usepackage{graphicx}

\def\3{2.8in}    
\def\2{2.5in}
\def\4{3.0in}

\begin{document}

\title{Superconductivity in the Bi/Ni bilayer}

\author{Sung-Po Chao}
\affiliation{Department of Physics, National Kaohsiung Normal University, Kaohsiung 82444, Taiwan, R.O.C.}



\begin{abstract}
Motivated by the recent observations of possible p-wave superconductivity in the bismuth-nickel (Bi/Ni) bilayer, we explore theoretically the possibilities of realizing p-wave superconductivity in this bilayer. We begin with 
a literature survey on this system and related materials which have similar superconducting transition temperature. From the survey the superconducting mechanism in this bilayer system is suggested to be phonon mediated type II superconductivity. A simple model is proposed to explain why the p-wave like Andreev reflection signals are likely to be observed in the surface probe, assuming the strong spin-orbit coupled surface state of Bi thin film is not completely destroyed by the formation of alloys.\\

\end{abstract}
\date{\today}
\maketitle
\section{Introduction}
The search for the possibility of realizing the simplest Non Abelian anyon, the Majorana zero modes\cite{Read2000, Kitaev2001} in the vortex state of topological superconductors, has been one of the hottest topics in the field of condensed matter. One way to achieve this topological superconductor is to search for superconductors with broken time reversal symmetry (TRS), implying the possible coexistence of ferromagnetism and superconductivity. Such coexistence has been reported in some heavy fermion materials, including UGe$_2$\cite{Saxena2000}, URhGe\cite{Aoki2001}, UCoGe\cite{Huy2007}, Sr$_2$RuO$_4$, etc. Among them Sr$_2$RuO$_4$ is perhaps the most extensively studied candidate of chiral p-wave superconductor, in which anomalous responses to external magnetic field and the appearance of half flux quantum has been reported. However, the lack of clear evidences for chiral edge current, a key signature for chiral superconductivity, has made the claim of Sr$_2$RuO$_4$ being a p-wave superconductor less conclusive.\\  

An alternative approach is to make artificial structures to realize such TRS broken superconductivity. Early proposals focus on the proximity effect between s-wave superconductor and ferromagnetic metal\cite{Efetov2001, Jonson2001} by artificially fabricating superconductor/ferromagnet(S/F) heterostructures. Numerous peculiar behaviors have been found in such systems such as spatial oscillation of electronic density of states, oscillatory superconducting transition temperature, and $\pi$ phase in Josephson junction to name a few. Recently with the emergence of topological insulators, it is proposed to use spin orbit coupled surface state of magnetic impurities doped topological insulators, or some semiconductors with strong spin orbit interactions placing in close proximity to a conventional superconductor combined with external magnetic field to achieve topological superconductivity. Zero bias anomaly\cite{Ng2009} in the tunneling differential conductance from the transport measurement is viewed as one of the important indications of obtaining Majorana zero modes. This zero bias anomaly has been experimentally observed in several candidate systems, including the point contact Andreev reflection spectra on the Bismuth surface of epitaxially grown Bismuth/Nickel (Bi/Ni) bilayer thin film observed by X. X. Gong et al.\cite{Jin2015} in 2015.\\ 

Bismuth/Nickel (Bi/Ni) bilayer film is viewed as one of the S/F heterostructures\cite{Moodera1990,Heiman2005}, but it is not a conventional S/F heterostructure as neither crystalline Bi nor Ni becomes superconducting above 1K at ambient pressure. Bulk crystalline Bi at ambient pressure enters into superconducting phase\cite{Prakash2017} at temperature below $0.53$mK due to its low carrier density. Making amorphous or polycrystalline Bi enhances the carrier density of state and the electron phonon interactions, brining up the superconducting transition temperature $T_c$ to $5\sim6$ K at ambient pressure. Similar enhancement is also achieved through placing single crystal under high pressure. Ni is a weak ferromagnetic metal, which shows no traces of superconductivity down to any measurable temperature. In 1990, it is found by J. S. Moodera and R. Meservey\cite{Moodera1990} that growing thin film of Bi on top of Ni thin film makes Bi side superconduct with optimal $T_c\simeq 4K$. In 2015, X. X. Gong et al.\cite{Jin2015} find zero bias anomaly which sustains even under high magnetic field in their transport measurement on the Bi/Ni thin film. \\  

In our model, we assume that the observed superconductivity in Bi/Ni is of conventional phonon mediated type and existing in the bulk of Bi thin film. This is supported by previous experimental reports\cite{Pratap2015,Chao2017,Liu2018} showing traces of alloyic Bi$_3$Ni formed throughout the Bi, even though this randomly distributed amount of alloys may not be significant enough to distort the X-ray diffraction images. That is, those diffusively formed random impurities do not significantly blur the rhombohedral structure in the Bi layer observed using transmission electron microscopy or X-ray diffraction. This alloyic Bi$_3$Ni has superconducting transition temperature around 4K and is a type II superconductor\cite{Awana2011}. Those properties explain why this Bi/Ni superconductivity could sustain with Ni thickness increasing up to around $1/5$ of the Bi thickness, and the reason for the maximal transition temperature in this bilayer is around 4K. The remaining puzzle is then the zero bias anomaly seen in the Andreev reflection signals\cite{Jin2015} observed on the Bi surface away from the Bi/Ni interface. We propose a simple theoretical model to explore the physical parameters regime that can realize this effective p-wave superconductivity on the Bi surface in this bilayer system. The mechanism is very similar to the effective p-wave superconductivity on the semiconductor surface with strong Rashba spin orbit coupling in close proximity with a conventional superconductor\cite{Sau2010,Jason2010,Kane2008}. We also summarize other possible mechanisms or explanations for seeing the zero bias anomaly in this bilayer Bi/Ni system.\\  

The rest of the paper is organized as follows. In section \ref{sect2} we briefly survey the literatures relevant to this Bi/Ni bilayer and summarize their claims. In section \ref{sect3} we propose a simple model to search for the physical parameters which can realize the possible time reversal broken $p\pm ip$ superconductivity in this Bi/Ni bilayer. Alternative explanations for the zero bias anomaly is also provided in the end of this section. In section \ref{sect4} we summarize our results, and suggest further experiments to explore this interesting Bi/Ni bilayer.

\section{Literature survey}\label{sect2}
Back in 1990 tunneling experiments on Bi/Ni bilayer done by J. S. Moodera and R. Meservey\cite{Moodera1990} shows that only Bi grown on top of Ni has superconductivity with $T_c\simeq 2\sim 4$K, while the reverse growth does not go into the superconducting phase. Simultaneous growth of Bi and Ni does not give superconductivity, and thus the alloyic Bi$_3$Ni superconductivity is ruled out. Based on this result they suspect the superconductivity is caused by the novel fcc structure, judged by the X-ray diffraction(XRD) patterns, grown on top of Ni.\\ 

However, their interpretation for the XRD data is controversial. The XRD data could also be explained by the common rhombohedral phase with the surface oriented along (110) direction instead of the novel fcc structure, as pointed out by J. A. van Hulst et al\cite{Jaeger93}. Recent experiments\cite{Jin2015,Zhou2017,Chao2017} with similar sample growth conditions show that the order of growth does not change its superconducting properties. The Bi surface orientation away from the Bi/Ni interface for thinner Bi is (110), while it changes to (111) for Bi thicker than 20 nm regardless of its order of growth. The reason for this discrepancy is not known, but suspected to be better control (better vacuum condition or lower substrate temperature) over the sample growth in the present day setup. \\

The tunneling transport and magnetic susceptibility measurements in Ref.~\onlinecite{Moodera1990} indicates the superconductivity in Bi/Ni is a strong to intermediate coupling ($2\Delta/k_BT_c\simeq 4$) type II s-wave superconductivity with upper critical field up to a few Tesla. Anisotropy in the critical field and tunneling measurement indicates this thin film superconductivity is not limited to the Bi/Ni interface but spreading out within the Bi layer. The normal state resistance of Bi/Ni is shown to be metallic (resistance drops with lowering of temperature) which is different from the insulating behavior seen in the pure Bi thin film\cite{Jin2015}. Ferromagnetism in the Ni layer is reduced in Bi/Ni compared with standalone Ni\cite{Moodera1990,Heiman2005,Chao2017}. In 2015, it is claimed to be p-wave like rather than s-wave superconductivity based on the Andreev reflection signal shown in the Ref.~\onlinecite{Jin2015}.\\
 
Artificially synthesized Bi$_3$Ni is shown to be a type II superconductor\cite{Awana2011,Zhao2018} with $T_c\simeq 4$K. The measured (bulk) upper critical field is in the order of $10^{-1}$T. Making Bi$_3$Ni in the form of thin film is expected to enhance its critical field. Since spectroscopy data\cite{Pratap2015,Chao2017} shows the formation of Bi$_3$Ni alloys, we suggest the superconductivity seen in the Bi/Ni bilayer is from the diffusively formed Bi$_3$Ni alloys. This viewpoint is also supported by the interesting experimental work done by L. Y. Liu et al.\cite{Liu2018}. In their work not only Bi$_3$Ni but also another type of alloy BiNi (with superconducting transition temperature $4.25$K) contributes to the superconductivity seen in the Bi/Ni bilayer. Due to the different growth methods (mainly pulsed laser deposition (PLD) in their work), the Ni ions have different spatial distributions in their work compared with others. Albeit with very similar transition temperature, those alloys have very different magneto-responses\cite{Liu2018} but the reason for superconductivity to happen in their samples is due to the formation of superconducting alloys. The remaining question then is that if we could also see the zero bias conductance peak in the point contact measurement, suggesting unconventional superconductivity in this Bi/Ni bilayer.\\

\underline{\textit{Update after posting this paper on the ArXiv:}}\\
 
Soon after this paper posted on the ArXiv, an experimental report on the study of superconductivity in the Bi/Ni bilayer is published by N. P. Armitage et. al.\cite{Armitage2019}. They use time domain THz spectroscopy to measure the low energy electrodynamic response of a Bi/Ni thin film. From their analysis the superconductivity is found to be fully gapped, and the superconductivity develops over the entire bilayer. Their experimental results are consistent with the s-wave bulk superconductivity in this bilayer system. \\

\section{Theoretical proposal for unconventional superconductivity}\label{sect3}
If the superconductivity in Bi/Ni bilayers observed in the Bi/Ni bilayer is due to the alloy formation, the observed superconductivity should be conventional phonon mediated s-wave superconductivity. We claim that, based on the theoretical model presented here, it is still possible to observe the two dimensional p-wave like superconductivity as seen in the Ref.~\onlinecite{Jin2015} on the Bi surface in our samples.\\

\begin{figure}
\centering
\includegraphics[width=0.5\textwidth]{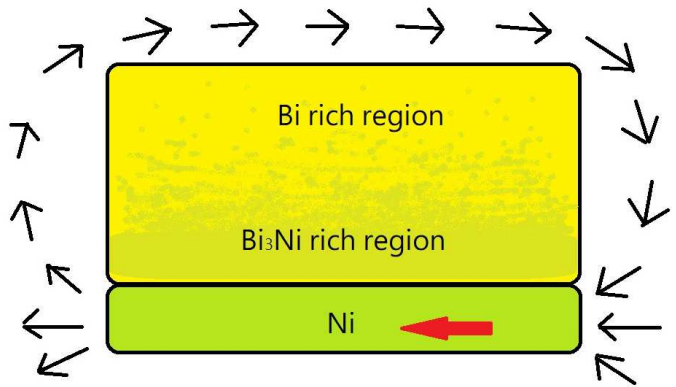}
\caption{\label{plot1} Sketch of the Bi/Ni bilayer. The magnetization of Ni, shown as a red arrow, is mostly in plane, and the magnetic field provided by this Ni layer on the Bi surface away from the interface is depicted by the black arrows. The field orientation at the Bi surface is also mostly in-plane. Alloys (mostly Bi$_3$Ni and few BiNi) is formed with higher concentration near the interface in the layer by layer epitaxial growth\cite{Pratap2015,Chao2017}, but the formation can vary with different growth techniques\cite{Liu2018}. At the top (Bi) surface of the figure, the spin orbit coupled surface state from Bi is assumed to be intact. The effective Hamiltonian for the Bi surface away from the interface is described in the Eq.(\ref{m1}) to Eq.(\ref{m4}).}
\end{figure}

 The basic idea is very similar to the proximity induced topological superconductivity using conventional s-wave superconductor in contact with a strong spin orbit interaction material under external magnetic field (or coupled with a ferromagnetic insulator)\cite{Sau2010,Jason2010}. Bi thin film is known to have robust metallic surface state\cite{Jin2012} and strong Rashba spin orbit interaction\cite{Hofmann2004,Hasegawa2006} on its surface. The alloyic Bi$_3$Ni provides the platform for conventional type II s-wave superconductivity. The required magnetic field\cite{Sau2010} is provided by the nickel thin film. This scenario is illustrated in the Fig.\ref{plot1}. Thus we have all the ingredients needed for realizing topological superconductivity in this Bi/Ni bilayer.\\
 
 Below we present our model Hamiltonian and the details of our theoretical results. The assumptions made in this model are that the spin orbit coupled surface state is not destroyed by the formation of a few randomly distributed alloys within the Bi layer, and the chemical potential of the sample is shifted to the region where topological superconductivity can be realized. The assumption of the spin orbit coupled surface state (not protected by band topology) on the Bi surface away from the interface is backed by the nice crystalline structure seen in the XRD and TEM\cite{Jin2015,Chao2017} in the Bi layer of the Bi/Ni bilayer, and the edge state property is not influenced by the local matrix properties away from the top surface in the tight binding model\cite{Aono2016}. However, the validity of existence of spin coupled surface state in the normal state should be checked by other surface probes such as Angle resolved photo emission spectroscopy(ARPES) or spin resolved scanning tunneling microscope(STM).

\subsection{Model Hamiltonian for $p\pm ip$ superconductivity on the surface of Bi}
Bi thin film has been shown to have robust metallic surface state, and the bulk state is changed from semi-metallic to insulating one as the thickness decreases\cite{Jin2012}. In forming the Ni/Bi interface, the smaller size of Ni allows Ni atoms to flow into the Bi layer, forming the superconducting alloy Bi$_3$Ni which has optimal critical temperature around $4K$. This alloy formation also serves as effective doping, leading to the shift of chemical potential. This is reflected in the normal state resistance seen in the Fig.1 of Ref.~\onlinecite{Jin2015} and similarly in the Ref.~\onlinecite{Aono2016}. Changes in the normal state charge carriers in the Bi/Ni bilayer compared with Bi thin film, using Hall bar measurements, also support this change in the chemical potential. At higher temperature the resistance goes up as rather than coming down as in the pure Bi thin film\cite{Jin2012}. This effective doping makes Bi/Ni bilayer metallic rather than insulator-like, which is the case for pure Bi thin film\cite{Jin2012}.\\

Suppose the formation of alloys is mostly nearby the Bi-Ni interface, the crystalline structure of bismunth away from the interface region will not be significantly modified. This claim is backed by the nice XRD data and trace of alloys seen in the experiments\cite{Pratap2015}, although the actual distributions of the alloys may depend on the details of the growth procedures\cite{Liu2018}. Under this assumption, we can treat the alloys as few random impurities in the region of bismunth away from the interface, and the effect of random impurities would only modifies the chemical potential without hampering the surface states with strong Rashba spin orbit coupling.\\
  
The effective Hamiltonian for surface state of Bi with surface oriented in the (111) direction\cite{Aono2016} in proximity with the bulk superconducting alloyic Bi/Ni thin film can be written as\cite{Jason2010} 
\bea\label{m1}
H&=&H_{Bi(s)}+H_{Z}+H_{Sc}\\\nn
H_{Bi(s)}&=&\int d^2 r \psi^{\dagger}\Big[-\left(\frac{\partial_x^2}{2m_x}+\frac{\partial_y^2}{2m_y}\right)-\mu-i(\alpha_x\sigma^x\partial_y\\\nn &&\quad \quad \quad-\alpha_y\sigma^y\partial_x)-i\alpha_z\sigma^z\Big((\partial_x+i\tilde{\beta}\partial_y)^3\\\label{m2} &&\quad \quad \quad +(\partial_x-i\tilde{\beta}\partial_y)^3\Big)\Big]\psi\\\label{m3}
H_{Z}&=&\int d^2 r \psi^{\dagger}\left(\vec{h}\cdot\vec{\sigma}\right)\psi\\\label{m4}
H_{Sc}&=&\int d^2 r\left(\Delta\psi^{\dagger}_{\uparrow}\psi^{\dagger}_{\downarrow}+\Delta^{\ast}\psi_{\downarrow}\psi_{\uparrow}\right)
\eea
The Hamiltonian for Bi surface $H_{Bi(s)}$ in the Eq.(\ref{m2}) describes the low energy dispersion of the bismuth up to cubic order near the $\Gamma$ point. This includes the usual quadratic kinetic energy, the linear in momentum Rashba spin orbit couplings due to broken inversion symmetry on the surface, and the cubic warping terms which results from the hexagonal lattice of bismuth. The $2\times 2$ Pauli matrices $\sigma_i$ act on the spin degree of freedom in $\psi(r)=\begin{pmatrix}\psi_{\uparrow}(r)\\ \psi_{\downarrow}(r)\end{pmatrix}$. This low energy effective Hamiltonian around $\Gamma$ point for (111) orientation is similar to that for (110) orientation\cite{Koroteev2008,Hirahara2007,Pascual2004}, which is seen for thinner sample as in the Ref~\onlinecite{Jin2015}, with some parameters change. Eq.(\ref{m3}) stands for the Zeeman field generated by the nickel thin film on the bismuth surface, and Eq.(\ref{m4}) is the proximity induced superconductivity on the surface state of bismuth thin film. The orbital term from the magnetic field is not included because the magnetic field generated from nickel film oriented in the in plane direction of the Bi/Ni bilayer. The magnitude of this in plane field decreases with increasing thickness of bismuth film (roughly proportional to inverse of the thickness to the third power, had we treated the nickel film as some bar magnet).\\

 The proximity induced superconductivity pairing amplitude $\Delta$ also decreases with increasing thickness of Bi (with Ni thickness fixed), as the parenting superconductor is formed by the alloyic Bi$_3$Ni whose formation is limited by the diffusive motion of nickel in the bismuth. Both of these factors contribute to the disappearance of $p\pm ip$ superconductivity in the bismuth surface as we increase the thickness of bismuth film. If we chose a thinner bismuth film, the magnetic field generated by the nickel could be large enough to kill the superconductivity of alloyic Bi$_3$Ni. Following these arguments we see that for a given thickness of nickel film there could be only a limited range of thickness of bismuth film giving rise to the superconductivity within the bulk of bismunth layer, which is consistent with the observations in the Ref.~\onlinecite{Jin2015}.\\

We further simplify Eq.(\ref{m2}) by rescaling $\partial_x\rightarrow(m_x/m_y)^{1/4}\partial_x$ and $\partial_y\rightarrow(m_y/m_x)^{1/4}\partial_y$. After this rescaling $H_{Bi(s)}$ becomes:
\bea\nn
H_{Bi(s)}&=&\int d^2 r \psi^{\dagger}\Big[-\frac{\nabla^2}{2m^*}-\mu-i\lambda_R(\sigma_x\partial_y-\gamma\sigma_y\partial_x)\\\label{bis}&+&i\lambda_D\sigma_z(\partial_x^3-3\beta\partial_x\partial_y^2)\Big]\psi 
\eea
 Here $m^*=\sqrt{m_xm_y}$, and the spin orbit coupling related parameters are $\lambda_D=2\alpha_z (m_x/m_y)^{3/4}$, $\beta=\tilde{\beta}^2 (m_x/m_y)$, $\gamma=(\alpha_y/\alpha_x)\sqrt{m_x/m_y}$, and $\lambda_R=\alpha_x(m_y/m_x)^{1/4}$. Eq.(\ref{bis}) is very similar to the low energy Hamiltonian describing the (110) quantum well mentioned in the Ref.~\onlinecite{Jason2010}, with the linear momentum dependent Dresselhaus term in the Ref.~\onlinecite{Jason2010} replaced by the cubic warping terms. Thus the physics leading to topological superconductivity with in plane magnetic field provided by nickel film here is basically the same as that of topological superconductivity formed by (110) quantum well with Rashba and Dresselhaus interactions with in plane magnetic field and in contact with a s-wave superconductor\cite{Jason2010}.\\

\begin{figure}
\centering
\includegraphics[width=0.5\textwidth]{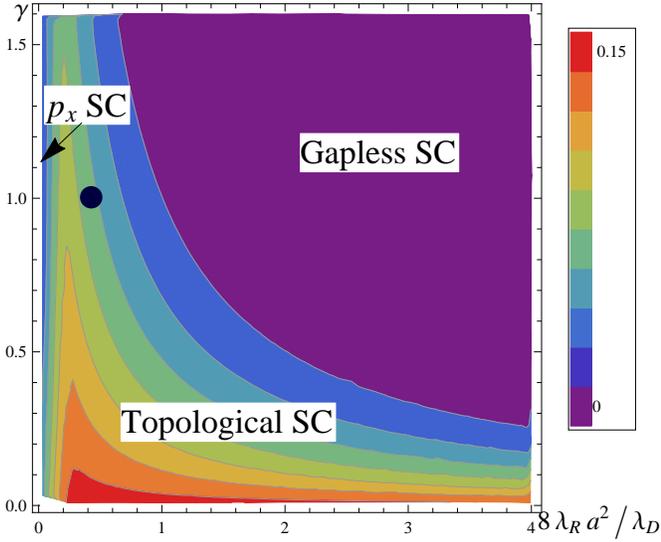}
\caption{\label{ph1} Phase diagram for gap magnitude $\mathcal{E}_g/\Delta$  as a function of anisotropic parameters $\gamma$ (with $\gamma=\beta$) and ratio between Rashba interaction $\lambda_R$ and warping terms induced Dresselhaus like interaction $\lambda_D$. Other fixed parameters are: $1/2m^{\ast}a^2=0.6$eV ,  lattice constant $a=4.53\AA$, chemical potential $\mu=0.9$eV, and effective Zeeman field from Ni layer around 2 meV. The dark blue dot corresponds to $\lambda_R/a=0.05$eV and $\lambda_D/a^3=0.8$eV, or $8\lambda_R a^2/\lambda_D=0.5$. The vertical axis corresponds to p-wave phase, similar to the phase diagram in the Ref.~\onlinecite{Jason2010}.}
\end{figure}

\begin{figure}
\centering
\includegraphics[width=0.5\textwidth]{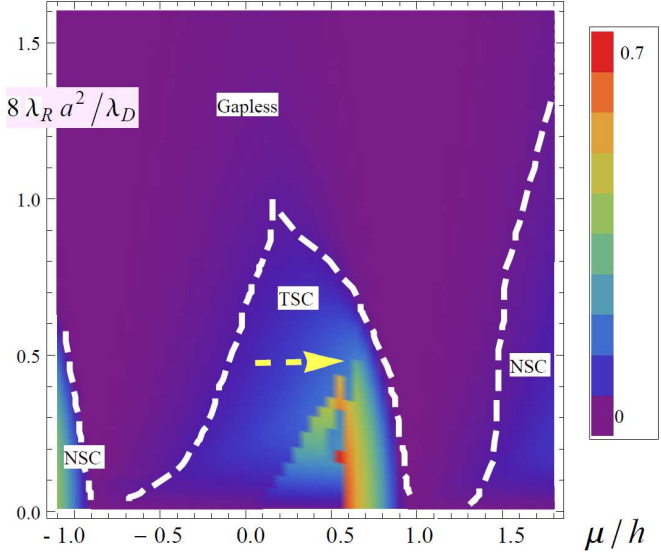}
\caption{\label{ph2} Phase diagram for gap magnitude $\mathcal{E}_g/\Delta$ as a function of the ratio between Rashba interaction $\lambda_R$ and warping terms induced Dresselhaus like interaction $\lambda_D$ and chemical potential $\mu$ normalized by the Zeeman field strength $h$. Other fixed parameters are: $\gamma=\beta=1$, $1/2m^{\ast}a^2=0.6$eV ,  lattice constant $a=4.53\AA$, $8\lambda_R a^2/\lambda_D=0.5$, and effective Zeeman field $h=2$ meV. The yellow arrow points to the maximal $p\pm ip$ order parameter magnitude which is around 0.4 meV in this calculation. NSC stands for normal superconducting state and TSC stands for topological superconducting state.}
\end{figure}

We rewrite the full Hamiltonian $H$ in momentum space and use diagonalized bases of $H_{Bi(s)}+H_{Z}$ by setting $\psi(\vec{k})=\begin{pmatrix}\psi_{\uparrow}(\vec{k}) \\\psi_{\downarrow}(\vec{k})\end{pmatrix}=\phi_-(\vec{k})\psi_-(\vec{k})+\phi_+(\vec{k})\psi_+(\vec{k})$. Here $\phi_{\pm}(\vec{k})$ represent some $2\times 1$ matrices, and $\psi_{\pm}(\vec{k})$ are the fermion annihilation operators for upper/lower bands. This is done in the same way as done in the Sau-Lutchyn-Tewari-Das Sarmar proposal for realizing topological superconductivity\cite{Sau2010}, which we summarize their results in the Appendix A. Here the explicit forms of $\phi_{\pm}(\vec{k})$ are not as illuminating as the case shown in the Appendix A, and we do not show their explicit forms. Following this bases change, we get:
\bea\nn
H&=&\int d^2 \vec{k} \Big[\left(\bar{\epsilon}_+(\vec{k})\psi^{\dagger}_+(\vec{k})\psi_+(\vec{k})+\bar{\epsilon}_-(\vec{k})\psi^{\dagger}_-(\vec{k})\psi_-(\vec{k})\right)\\\nn&+&\Big(\Delta_{+-}(\vec{k})\psi^{\dagger}_+(\vec{k})\psi^{\dagger}_-(-\vec{k})
+\Delta_{++}(\vec{k})\psi^{\dagger}_+(\vec{k})\psi^{\dagger}_+(-\vec{k})\\\label{fullH}&+&\Delta_{--}(\vec{k})\psi^{\dagger}_-(\vec{k})\psi^{\dagger}_-(-\vec{k})+h.c.\Big)\Big]
\eea
The upper/lower band energies $\bar{\epsilon}_{\pm}(\vec{k})$ are given by 
\bw\bea\nn
&&\bar{\epsilon}_{\pm}(\vec{k})=\frac{k^2}{2m^*}-\mu\pm\delta\epsilon(\vec{k})\quad,\\\label{eppm}
&&\delta\epsilon(\vec{k})=\sqrt{(\gamma\lambda_Rk_x-h_y)^2+(\lambda_D(k_x^3-3\beta k_xk_y^2)+h_z)^2+(\lambda_Rk_y+h_x)^2}.
\eea\ew
Using this band bases, the $s$-wave like interband pairing strength $|\Delta_{+-}(\vec{k})|$ and $p\pm ip$-wave like intraband pairing ($|\Delta_{++}(\vec{k})|$ or $|\Delta_{--}(\vec{k})|$) are expressed as:
\bw
\bea
&&|\Delta_{+-}(\vec{k})|^2=\frac{\Delta^2}{2}\left[1-\frac{\lambda_D^2(k_x^3-3\beta k_xk_y^2)^2+\gamma^2\lambda_R^2k_x^2+\lambda_R^2k_y^2-(h_x^2+h_y^2+h_z^2)}{\delta\epsilon(\vec{k})\delta\epsilon(-\vec{k})}\right],\\\nn
&&|\Delta_{++}(\vec{k})|^2=|\Delta_{--}(\vec{k})|^2=\frac{\Delta^2}{8}\left[1+\frac{\lambda_D^2(k_x^3-3\beta k_xk_y^2)^2+\gamma^2\lambda_R^2k_x^2+\lambda_R^2k_y^2-(h_x^2+h_y^2+h_z^2)}{\delta\epsilon(\vec{k})\delta\epsilon(-\vec{k})}\right].
\eea
\ew
Solving the full Bogoliubov-de Gennes Hamiltonian obtained from Eq.(\ref{fullH}) with uniform $\Delta$, we get
\bea\nn
&&E_{\pm}(\vec{k})^2=4|\Delta_{++}(\vec{k})|^2+|\Delta_{+-}(\vec{k})|^2+\frac{\bar{\epsilon}_+(\vec{k})^2+\bar{\epsilon}_-(\vec{k})^2}{2}\\ &&\pm |\bar{\epsilon}_+(\vec{k})-\bar{\epsilon}_-(\vec{k})|\sqrt{|\Delta_{+-}(\vec{k})|^2+\Big(\frac{\bar{\epsilon}_+(\vec{k}) +\bar{\epsilon}_-(\vec{k})}{2}\Big)^2}
\eea
We concentrate on the lower branch $E_-(\vec{k})$ , assuming the chemical potential of the bismunth surface state is lowered to be around $\frac{k^2}{2m^\ast}-\delta\epsilon(\vec{k})$, with $\vec{k}$ around the $\Gamma$ point in the momentum space. The lowering of the chemical potential could come from the effective doping due to the formation of random alloyic Bi$_3$Ni impurities within the bulk of bismunth. This is suggested by the different temperature dependence of the resistance in the normal state of pure bismunth\cite{Jin2012} and Bi/Ni\cite{Jin2015} thin film. The minimum of $E_-(\vec{k})$ around $\Gamma$ point determines the superconducting gap, denoted as $\mathcal{E}_g$ in the Fig.\ref{ph1} and Fig.\ref{ph2}, of the surface state. The change in $\mathcal{E}_g$ computed numerically is used to explore the stability conditions of various topological and non-topological phases evaluated at zero temperature as shown in the Fig.\ref{ph1} and Fig.\ref{ph2}. Finite temperature phase diagram can be done by constructing its Helmholtz free energy. As the goal here is to find the maximal proximity induced topological superconducting gap in the model Hamiltonian, we adhere to the zero temperature formulation throughout this paper.\\

The phase diagram shown in the Fig.\ref{ph1} is $\mathcal{E}_g$ evaluated as a function of anisotropy parameter $\gamma$ (setting $\gamma=\beta$ to simplify the phase diagram) and dimensionless ratio between Rashba interaction strength $\lambda_R$ and warping terms induced Dresselhaus like interaction strength $\lambda_D$ at fixed chemical potential $\mu$ and Zeeman energy $|\vec{h}|\equiv h$ (with Zeeman field chosen to be along in-plane y-axis in both figures) evaluated at experimental relevant values discussed in the section \ref{mpr}. In Fig.\ref{ph1} the chemical potential $\mu$ is chosen to place the Fermi level crossing only the lower band, with energy dispersion $\bar{\epsilon}_-(\vec{k})$. The separation in energy from the upper band to lower one is mainly determined by $h$, which is chosen to be larger than the proximity induced superconducting gap magnitude $\Delta$.\\ 

For $\vec{h}=h_y$ considered here, the nonzero $\gamma$ lifts the $k_x\rightarrow -k_x$ symmetry of the $\Delta=0$ bands as can be seen from Eq.(\ref{eppm}). This suppresses the superconductivity since the pairing state involves $\vec{k}$ and $-\vec{k}$. Smaller $\lambda_R$ partially offsets this effect, and thus the smaller superconducting pairing gap (with larger $\gamma$ and $\lambda_R$) and the region termed "gapless superconductor" is located in the upper right corner of the Fig.\ref{ph1}. We shall emphasize that the proximity effect generates not only interband s-wave like pairing ($\Delta_{+-}$), but also intraband $p\pm ip$ like pairing ($\Delta_{++}$/$\Delta_{--}$) in the upper/lower band. The gapless superconducting region (where $\mathcal{E}_g=0$) does not mean that $\Delta_{--}$ or $\Delta_{+-}$ is zero, but rather a region of transition from topological nontrivial superconductor to topological trivial superconductor as in the cases of Sau et. al.'s model\cite{Sau2010} and Jason Alicea's model\cite{Jason2010}. For
$\lambda_R\rightarrow 0$ but finite $\lambda_D$ (region close to y-axis in the Fig.\ref{ph1}) the dominant spin orbit coupling for $\Delta_{--}$ is the warping induced Dresselhaus like spin orbit coupling. Right at the y-axis, the induced topological superconductivity pairing symmetry is given by $k_x(k_x^2-3\beta k_y^2)\simeq -k_x|\vec{k}|^2$ (for $\beta\simeq 1$, $|k_x|\simeq|k_y|$ in an isotropic sample; $|\vec{k}|^2=k_x^2+k_y^2$) behaving like $p_x$ superconductor nearby $\Gamma$ point. Thus we mark that region as a "$p_x$ superconductor.\\

The topological phase transition is also present when we try to move the chemical potential away from the lower band. An naive guess for the phase boundary would be $|\mu|\le |h|$ as the "topological gap" is protected by the Zeeman field here (the actual topological region would be smaller as $\Delta$ is finite). Thus the phase boundary, or region named gapless superconductor, exists between the "normal superconductor (NSC)" and "topological superconductor (TSC)" as we change the chemical potential while fixing other parameters as shown in the Fig.\ref{ph2}.   

\subsection{Model parameters relevant to the known experimental results}\label{mpr} 
We use the model parameters $1/2m^{\ast}a^2=0.6$eV , $\lambda_R/a=0.05$eV, $\lambda_D/a^3=0.8$eV, lattice constant $a=4.53\AA$, and chemical potential $\mu=0.9$eV to fit the hexagonal Fermi surface of the pristine bismunth thin film\cite{Aono2016} around the $\Gamma$ point. With the addition of nickel layer, we lower the chemical potential to zero and add effective Zeeman field of magnitude around $2$meV. The upper bound of Zeeman field is estimated from the in plane upper critical field\cite{Pratap2015} of Bi$_3$Ni (with thickness around one tenth of magnetic penetration depth) and large gyromagnetic ratio ($g\simeq 33$) of Bi thin film, which gives $10$ meV. To keep the alloyic superconductivity from Bi$_3$Ni as intact as possible, we choose the Zeeman field $h$ generated from the nickel layer to be $2$meV. The superconducting gap magnitude $\Delta$ from the bulk Bi is estimated to be $0.9$meV ,using $2\Delta/k_BT_c=4.5$ and $T_c=4$K measured by tunneling experiment in a similar setup\cite{Moodera1990}.\\

With aforementioned parameters and assuming the film is uniform (with dimensionless anisotropic parameters $\gamma=1$, $\beta=1$), the surface state of Bi/Ni is then described by $p+ip$ topological superconductivity with superconducting gap magnitude around $0.08$meV or $0.09\Delta$(dark blue dot in the Fig.\ref{ph1}). Further lifting up of chemical potential (say, by around $1$ meV) with other parameters fixed enhances the gap magnitude up to $0.4$ meV as shown in the Fig.\ref{ph2}. This enhancement is attributed to the enlargement of density of state with the rising of chemical potential. Further increasing of chemical potential results in change from topological superconductivity to topologically trivial one. Making the film anisotropic also leads to a larger gap, although the effect is less significant compared with the shift of chemical potential.\\ 

Fitting using generalized Blonder, Tinkham, and Klapwijk (BTK) formula\cite{Tanaka2007} with the observed zero bias peak\cite{Jin2015} gives superconducting gap around $0.6$ to $1.1$ meV (depending on the choice of fitting range of bias voltage). This estimated gap magnitude is almost the same as that from the bulk superconductivity ($\sim 0.9$meV) with critical temperature around 4K. Our numerical results for largest proximity induced gap magnitude is around $0.4$meV. This factor of two discrepancy could come from the thermal broadening or multichannel tunnelings due to finite size of the point contact. Further reducing the measurement temperature and choosing a better contact could possibly resolve this issue. 

\subsection{Anisotropic point contact Andreev reflection}
\begin{widetext}
\begin{figure}
\centering
\includegraphics[width=1\textwidth]{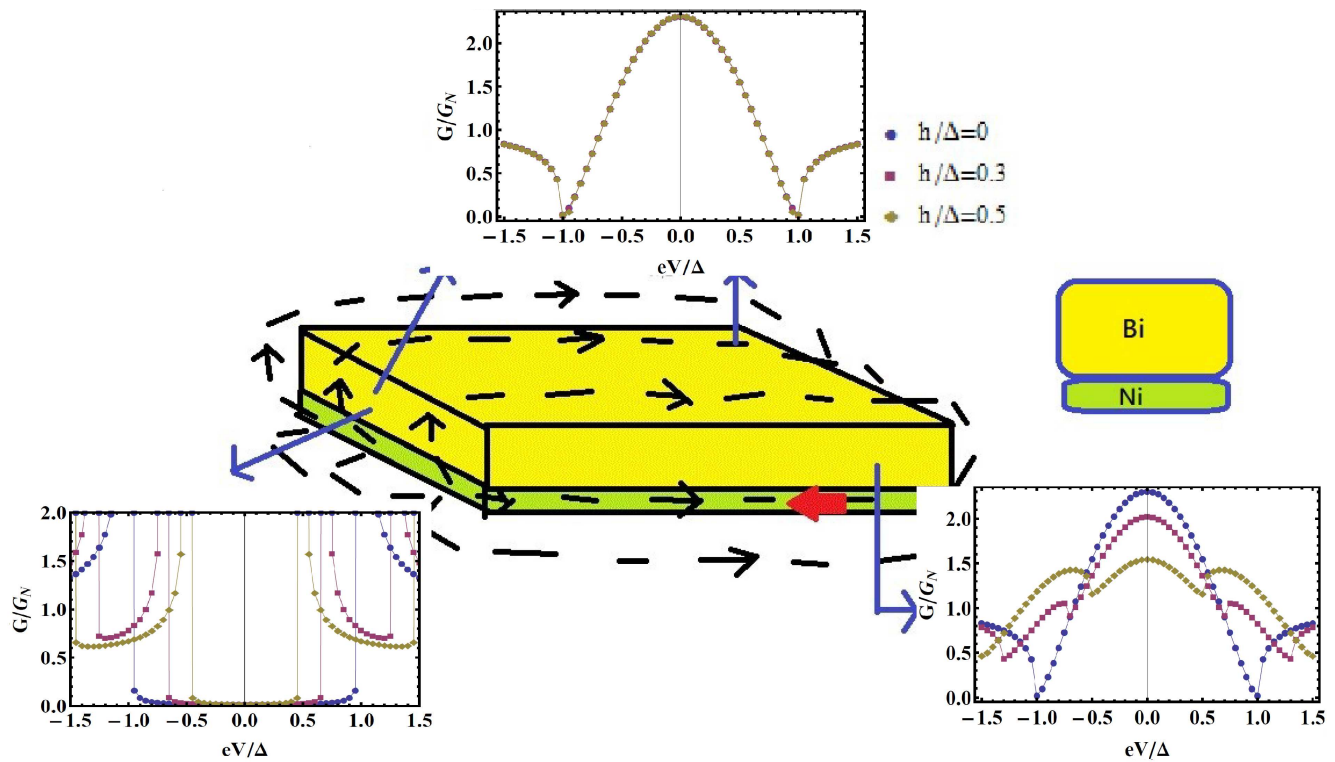}
\parbox{\textwidth}{\caption{\label{ph3} Schematic plots for the anisotropic conductance from the point contact Andreev reflection measurement. Formula used in the evaluation of normalized conductance shown in the three sub-figures is taken from the Ref.~\onlinecite{Tanaka2007}, with $Z=\frac{2mU}{\hbar^2 k_F}=10$ ($U$ being the normal-superconductor interface potential) as in the Fig.1 of the Ref.~\onlinecite{Tanaka2007}. Different colors in the sub-figures means normalized conductance (vertical axis: $G/G_N$ where $G_N$ stands for normal state conductance) v.s. bias voltage (horizontal axis: $eV/\Delta$ where $V$ is the bias voltage) under different local magnetic fields.}}
\end{figure}
\end{widetext}

Another interesting perspective of this Bi/Ni bilayer is that the differential conductance signal from the point contact shows different results from different sides\cite{Jintalk2016,Chientalk2018}. Effective equal spin $p\pm ip$ pairing is expected to be strongest only on the surface away from the Bi/Ni interface. For the side surfaces the conductance shape could show s-wave like or p-wave like structures. For rectangular shape of Bi/Ni bilayer, the side with longer length is supposed to be influenced by similar magnetic field strength as the top surface. The side with shorter length experiences stronger magnetic field compared with that of the top surface, but with more anisotropy in field distribution as illustrated in the cartoon picture of Fig.\ref{ph3}. This anisotropy leads to smaller effective in plane field compared with the other two side surfaces. Since the side surface Rashba terms are much weaker compared with that of the surface in parallel with the Bi/Ni interface, the side with large magnetic field is likely to show $p$-wave behavior as suggested in Fig.\ref{ph1} with $\lambda_R=0$. It could also shows opposite spin pairing\cite{Tanaka2007} in the triplet state which would be sensitive to the external magnetic field probes. For the side surface with weaker magnetic field, the chemical potential $\mu$ could be larger than the Zeeman field $h$, leading to $s$-wave like topologically trivial superconductivity as shown in right-hand side of the phase diagram in the Fig.\ref{ph2}.\\

\subsection{Competing models and other possibilities}
Recent optical measurements of the polar Kerr effect supports the spontaneous time reversal symmetry breaking on the Bi surface in concurrence with the onset of superconductivity in this Bi/Ni bilayer\cite{Xia2017}. This experimental results are consistent with the time reversal broken $p\pm ip$ paring gap presented in our theoretical model. An alternative theoretical explanation for the same Kerr effect results is presented in Ref.~\onlinecite{Xia2017}, where the superconductivity is thought to be occurring only on the Bi surface away from the Bi/Ni interface. Based on symmetry requirements for two dimensional noncentrosymmetric crystalline superconductor\cite{Samokhin2015}, the authors in Ref.~\onlinecite{Xia2017} concluded that the time reversal broken superconducting state should be of $d\pm id$ instead of $p\pm ip$ as suggested in our scenario. The mechanism behind this $d\pm id$ superconductivity is the magnetic fluctuations induced by the Ni layer.\\

The key difference between this $d\pm id$ proposal and ours is that the superconductivity considered in our model is rooted from the bulk of Bi, not just on the surface away from the Bi/Ni interface. It is true that with the decrease of Bi thickness the bulk of Bi tends to be a normal insulator with metallic surface state\cite{Jin2012} (with the exception of few bilayers of Bi which could be topological insulator\cite{Matsuda2016} or single layer of Bi as two dimensional topological insulator\cite{Wu2011}). However, by placing Bi on top of Ni thin film, we see the whole Bi/Ni normal state behaves like an usual metal rather than an insulator. This leads us to believe that, in all the Bi/Ni samples we see, there exist effective doping of charges which increase the electronic density of state nearby the Fermi level. Also all the observed transport and magnetic properties, other than the surface probes such as the point contact Andreev reflection, in the superconducting state is consistent with the usual type II s-wave superconductor. Another experimental support is that we do not see any sign of superconductivity in the Bi/Fe or Bi/Co samples\cite{Chao2017}, which should have similar superconducting behavior if the surface $d\pm id$ superconductivity were induced by magnetic fluctuations.\\     

It is also possible that the observed p-wave like signal in the point contact measurement is from the bulk of Bi/Ni bilayer\cite{Chientalk2018} instead of signals from the surface. It is found by T. Herrmannsd$\ddot{o}$rfer et. al.\cite{Ruck2011} that nano-structures of Bi$_3$Ni (submicrometer-sized particles and quasi-one-dimensional nanoscaled strains) also shows coexistence of superconductivity and ferromagnetism with onset superconducitng transition temperature around $5.2K$. This kind of nano-structured Bi$_3$Ni could also be formed during the epitaxial growth of Bi/Ni bilayer and becomes the source of the bulk p-wave superconductivity, although the mechanism behind it remains elusive. Whether these nanostructured Bi$_3$Ni could form well-oriented domains as suggested by the anisotropy measurement\cite{Jintalk2016,Chientalk2018} during the epitaxial growth is yet another puzzle to be clarified.\\     

Another possibility for seeing magnetic field independent zero bias conductance peak in the point contact Andreev reflection measurement is that the point contact is not in the Sharvin ballistic limit\cite{Greene2010}. This has been seen in some of the multibands iron-pnictide superconductors with s$\pm$ pairing symmetry. In the polycrystalline iron-pnictide it is also found\cite{Greene2010} the coexistence of randomly distributed ferromagnetic and superconducting domains. Both the field independent zero bias conductance peak and the existence of ferromagnetic domains could possibly explain the experimental results from the point contact and magneto-Kerr effect measurements found in the Bi/Ni samples. This less exciting possibility can be ruled out, if the superconducting site found in the point contact measurement were the same as the ferromagnetic region found in the Kerr effect measurement. Multiple Andreev reflections\cite{Bose2017} is yet another possibility, although the experimental results from Ref.~\onlinecite{Jin2015} suggests this is less likely to be the case.

\section{Summary and further experimental suggestions}\label{sect4}
 We propose a simple model, utilizing the strong spin orbit coupling nature of Bi and the effective doping coming from the alloy formation in the Bi/Ni bilayer, to suggest the possible existence of proximity induced time reversal broken $p\pm ip$ superconductivity on the Bi surface away from the Bi/Ni interface. The physics behind it is the same as the effective $p\pm ip$ superconductivity made by conventional superconductor combined with a semiconductor with strong spin orbit couplings under some external magnetic field\cite{Sau2010,Jason2010}. The key difference here is the Rashba spin orbit term is supplemented by a cubic spin orbit coupling, and the external magnetic field is provided by the ferromagnetic Ni layer. By mapping out the phase diagram with experimentally relevant parameters, we also explain the anisotropic Andreev reflection signals probed on different Bi surface\cite{Jintalk2016,Chientalk2018}. This $p\pm ip$ scenario is also consistent with the recent magneto-optical Kerr effect and magnetic measurements\cite{Xia2017,Zhou2017}, although other possibilities such as bulk p-wave superconductivity induced by nanostructured Bi$_3$Ni\cite{Ruck2011}, Multiple Andreev reflections\cite{Bose2017}, or point contact in the diffusive regime\cite{Greene2010} should also be considered. Since the alloy formations
would vary with growth methods, the phase diagrams mentioned in our simple model for the actual bilayer system is surely more complicated. However, we think the main physics that the topological nontrivial superconductivity is induced through proximity effect on the surface of Bi should be still the same, as long as the surface state of Bi away from the interface is not destroyed after the formation of those alloys from interface diffusions.\\

 To truly confirm whether our proposed scheme is correct or not, further surface probes such as Angle resolved photo emission spectroscopy or low temperature scanning tunneling microscope is needed to check the normal and superconducting state of electronic structures of Bi layer after forming the Bi/Ni bilayer. For sufficient thick Bi layer, the nickel from the interface diffusion shall not reach to the Bi surface away from the interface. The size of the superconducting gap from the point contact Andreev reflection measurement shall become smaller with the increasing bismuth thickness.\\ 
 
Another possible mechanism for inducing time reversal broken superconductivity is through the magnetic fluctuations from the Ni layer as mentioned in the Ref.~\onlinecite{Xia2017}. We tend to exclude this scenario based on the lack of superconductivity in the Bi/Fe and Bi/Co bilayers. Had it indeed been able to achieve effective $p\pm ip$ superconductivity on the Bi surface, we may adjust the sample makings processes to have the largest superconducting gap magnitude, using aforementioned phase diagrams, and make Majorana zero modes in its vortex state. Even if it were not the cases (say, with zero bias anomaly seen similar to some of the iron-pnictide superconductors), a further look at the Fulde Ferrell Larkin Ovchinnikov (FFLO) physics on the Ni side\cite{Heiman2005} is also an interesting topic in its own right. A systematic study of bilayer formed by metallic/semi-conducting thin film with strong spin orbit couplings and ferromagnetic metal/insulator layer could also possibly leads to the platform for effective $p\pm ip$ or even more exotic superconductors yet to be explored.\\ 

\acknowledgements
 I acknowledge useful discussions with Piers Coleman in the initial stage of this work, and various discussions with friends from Institute of Physics, Academia Sinica during the writeup of this work. I also thank the financial support from Taiwan's MOST (No.106-2112-M-017-002-MY3), NSF funding from U.S.A. for the Aspen center for physics 2017 winter conference, and Academia Sinica during my winter and summer visits in 2016-2017.\\

\appendix
\section{Overview of Sau-Lutchyn-Tewari-Das Sarmar proposal}
The mechanism for generating $p\pm ip$ superconductivity in this paper follows the idea pioneered by Sau et. al. in the Ref.~\onlinecite{Sau2010}, of utilizing semiconductors with strong spin orbit coupling under external magnetic field and placed in close contact with a conventional superconductor. We summarize their main formulations and results following the review article\cite{Jason2012} and paper\cite{Jason2010} by Jason Alicea.\\

The proposed setup for realizing $p\pm ip$ superconductivity is to use a semiconductor quantum well (a quasi two dimensional electron gas (2DEG)) with strong Rashba spin orbit coupling placed in between a conventional s-wave superconductor and a ferromagnetic insulator. The magnetization direction of the ferromagnetic insulator is pointing perpendicular to the plane of the two dimensional electron gas formed in the quantum well. This stacking order (s wave superonductor-2DEG-Ferromagnetic insulator) is different from the scenario for Bi/Ni bilayer (2DEG-s wave superonductor-Ferromagnetic metal) we mentioned in the main text. The direction of magnetization and the spin orbit coupling terms for generating  topological superconductivity is slightly different in our proposal.
In the following, we introduce term by term the low energy Hamiltonian describing the setup by Sau at. al.\cite{Sau2010}, starting with the 2DEG with Rashba spin orbit coupling part.\\ 

Up to quadratic order in momentum, the relevant Hamiltonian for the electrons in the quantum well is:
\bea\label{A1}
H_o=\int d^2 r\psi^{\dagger}\left[-\frac{\nabla^2}{2m}-\mu-i\alpha(\sigma^x\partial_y-\sigma^y\partial_x)\right]\psi
\eea
 where $m$ is the effective mass, $\mu$ is the chemical potential, $\alpha$ is the Rashba spin-orbit coupling strength, and $\sigma^i$ are Pauli matrices acting on the spin degree of freedom in $\psi$. At small enough momentum (around $\Gamma$ point in the reciprocal lattice) the Rashba term in the Eq.(\ref{A1}) gives spin-orbit coupled band similar to the electrons of the two dimensional surface state of a three dimensional topological insulator. The emergence of $p\pm ip$ superconductivity is closely related to this term.\\ 

Next the coupling with a ferromagnetic insulator with magnetization pointing out of plane is assumed to induce a Zeeman interaction:
\bea\label{A2}
H_Z=\int d^2 r\psi^{\dagger}[h_z\sigma^z]\psi
\eea 
Under the assumption that the influence from the ferromagnetic insulator were primarily due to exchange interaction, this Zeeman interaction is the dominant term. For $|h_z|>|\mu|$ the electrons occupy only the lower band and exhibit a single Fermi surface. To see this we may diagonalize $H_0+H_Z$ in momentum space by writing
\bea
\psi(\vec{k})=\phi_-(\vec{k})\psi_-(\vec{k})+\phi_+(\vec{k})\psi_+(\vec{k})
\eea
 where $\psi_{\pm}(\vec{k})$ are the fermion annihilation operators for upper/lower bands, and $\phi_{\pm}(\vec{k})$ are the corresponding wave functions taking the following form:
\bea
\phi_+(\vec{k})&=&\left(\begin{array}{c} A_{\uparrow}(\vec{k}) \\ A_{\downarrow}(\vec{k})\frac{ik_x-k_y}{|\vec{k}|} \end{array}\right) ,\\
\phi_-(\vec{k})&=&\left(\begin{array}{c} B_{\uparrow}(\vec{k})\frac{ik_x+k_y}{|\vec{k}|} \\ B_{\downarrow}(\vec{k}) \end{array}\right) .
\eea
The expression for $A_{\uparrow,\downarrow}$ and $B_{\uparrow,\downarrow}$ are not that enlightening\cite{Jason2010}, but the combination of them as shown below are more meaningful:
\bea\label{a6}
&&f_p(\vec{k})\equiv A_{\uparrow}A_{\downarrow}=B_{\uparrow}B_{\downarrow}=\frac{-\alpha |\vec{k}|}{2\sqrt{h_z^2+\alpha^2 |\vec{k}|^2}} ,\\\label{a7}
&&f_s(\vec{k})\equiv A_{\uparrow}B_{\downarrow}-B_{\uparrow}A_{\downarrow}=\frac{h_z}{\sqrt{h_z^2+\alpha^2 |\vec{k}|^2}} . 
\eea
Eq.(\ref{a6}) and Eq.(\ref{a7}) are useful for identifying the formation of $p\pm ip$ and $s$-wave superconductivity once the proximity induced pairing is introduced. In this band bases, $H_0+H_Z$ becomes:
\bea\nn
H_0+H_Z=\int d\vec{k}\left[\epsilon_+(\vec{k})\psi_{+}^{\dagger}(\vec{k})\psi_+(\vec{k})+\epsilon_-(\vec{k})\psi_{-}^{\dagger}(\vec{k})\psi_-(\vec{k})\right]
\eea
with energies
\bea\nn
\epsilon_{\pm}(\vec{k})=\frac{|\vec{k}|^2}{2m}-\mu\pm\sqrt{h_z^2+\alpha^2|\vec{k}|^2}.
\eea
Making this 2DEG in contact with a s-wave superconductor introduces a pairing term via the proximity effect. The full Hamiltonian is given by:
\bea\label{fH}
H=H_0+H_Z+H_{Sc}
\eea
 with
\bea\label{asc} 
H_{Sc}=\int d^2 r\left[\Delta \psi_{\uparrow}^{\dagger}\psi_{\downarrow}^{\dagger}+\Delta^{\ast}\psi_{\downarrow}\psi_{\uparrow}\right].
\eea
We choose $\Delta=\Delta^{\ast}$ in this s-wave pairing. Rewriting $H_{Sc}$ in terms of band bases $\psi_{\pm}$ in momentum space we get:
\bea\nn
&&H_{Sc}=\int d\vec{k} \Big[\Delta_{+-}(\vec{k})\psi^{\dagger}_+(\vec{k})\psi^{\dagger}_-(-\vec{k})+\Delta_{++}(\vec{k})\\\nn
&&\times\psi^{\dagger}_+(\vec{k})\psi^{\dagger}_+(-\vec{k})
+\Delta_{--}(\vec{k})\psi^{\dagger}_-(\vec{k})\psi^{\dagger}_-(-\vec{k})+h.c.\Big]
\eea
with
\bea
&&\Delta_{+-}(\vec{k})=f_s(\vec{k})\Delta,\\
&&\Delta_{++}(\vec{k})=f_p(\vec{k})\left(\frac{k_y+ik_x}{|\vec{k}|}\right)\Delta,\\
&&\Delta_{--}(\vec{k})=f_p(\vec{k})\left(\frac{k_y-ik_x}{|\vec{k}|}\right)\Delta.
\eea
In this band basis we see that Eq.(\ref{asc}) generates both interband s-wave paring $\Delta_{+-}$ and intraband $p\pm ip$ pairing for upper/lower bands. This mixing pairings is due to spin-momentum locking from the Rashba spin orbit coupling. For proximity effect induced pairing amplitude $\Delta$ much smaller than the band gap induced by the Zeeman field (i.e. $|h_z-\mu|\gg \Delta$) with $\mu$ crossing the lower band, we can ignore the upper band and project $\psi_+$ away from the effective Hamiltonian. Then the remaining effective Hamiltonian maps onto that of spinless $p-ip$ pairing, an example of topological superconductor. Notice that even in this topological superconducting regime the interband s-wave paring $\Delta_{+-}$ is still nonzero around $\Gamma$ point in reciprocal space. The reason for projecting interband pairing away is purely due to the energetic assumption that $|h_z-\mu|\gg \Delta$.\\

To explore the stability conditions at zero temperature for this topological superconducting phase, we solve the full Hamiltonian Eq.(\ref{fH}) and obtain:
\bea\nn
&&E_{\pm}^2=4|\Delta_{++}|^2+\Delta_{+-}^2+\frac{\epsilon_+^2+\epsilon_-^2}{2}\\
&&\pm|\epsilon_+-\epsilon_-|\sqrt{\Delta_{+-}^2+\frac{(\epsilon_+-\epsilon_-)^2}{4}}.
\eea
 For the lower band energy eigenvalue $E_-(\vec{k})$ the superconducting pairing amplitude is obtained at the Fermi momentum $k_F$ (obtained from $\epsilon_-(\vec{k}_F)=0$). The smallest $E_-(\vec{k})$ band gap is around $\vec{k}=0$ (the $\Gamma$ point in reciprocal space). This gap closing ($E_-(\vec{k})=0$ at some $\vec{k}$ close to $\Gamma$ point, named "gapless superconducting" region in the Fig.\ref{ph1},\ref{ph2} of the main text) marks the transition from topological superconducting phase (absolute value of Chern number equals to one) to trivial (normal) superconducting phase (Chern number equals to zero)\cite{Jason2012}. 
 
\bibliographystyle{apsrev4-1}
\bibliography{BiNiSc}

\begin{thebibliography}{41}%
\makeatletter
\providecommand \@ifxundefined [1]{%
 \@ifx{#1\undefined}
}%
\providecommand \@ifnum [1]{%
 \ifnum #1\expandafter \@firstoftwo
 \else \expandafter \@secondoftwo
 \fi
}%
\providecommand \@ifx [1]{%
 \ifx #1\expandafter \@firstoftwo
 \else \expandafter \@secondoftwo
 \fi
}%
\providecommand \natexlab [1]{#1}%
\providecommand \enquote  [1]{``#1''}%
\providecommand \bibnamefont  [1]{#1}%
\providecommand \bibfnamefont [1]{#1}%
\providecommand \citenamefont [1]{#1}%
\providecommand \href@noop [0]{\@secondoftwo}%
\providecommand \href [0]{\begingroup \@sanitize@url \@href}%
\providecommand \@href[1]{\@@startlink{#1}\@@href}%
\providecommand \@@href[1]{\endgroup#1\@@endlink}%
\providecommand \@sanitize@url [0]{\catcode `\\12\catcode `\$12\catcode
  `\&12\catcode `\#12\catcode `\^12\catcode `\_12\catcode `\%12\relax}%
\providecommand \@@startlink[1]{}%
\providecommand \@@endlink[0]{}%
\providecommand \url  [0]{\begingroup\@sanitize@url \@url }%
\providecommand \@url [1]{\endgroup\@href {#1}{\urlprefix }}%
\providecommand \urlprefix  [0]{URL }%
\providecommand \Eprint [0]{\href }%
\providecommand \doibase [0]{http://dx.doi.org/}%
\providecommand \selectlanguage [0]{\@gobble}%
\providecommand \bibinfo  [0]{\@secondoftwo}%
\providecommand \bibfield  [0]{\@secondoftwo}%
\providecommand \translation [1]{[#1]}%
\providecommand \BibitemOpen [0]{}%
\providecommand \bibitemStop [0]{}%
\providecommand \bibitemNoStop [0]{.\EOS\space}%
\providecommand \EOS [0]{\spacefactor3000\relax}%
\providecommand \BibitemShut  [1]{\csname bibitem#1\endcsname}%
\let\auto@bib@innerbib\@empty
\bibitem [{\citenamefont {Read}\ and\ \citenamefont {Green}(2000)}]{Read2000}%
  \BibitemOpen
  \bibfield  {author} {\bibinfo {author} {\bibfnamefont {N.}~\bibnamefont
  {Read}}\ and\ \bibinfo {author} {\bibfnamefont {D.}~\bibnamefont {Green}},\
  }\href@noop {} {\bibfield  {journal} {\bibinfo  {journal} {Phys. Rev. B}\
  }\textbf {\bibinfo {volume} {61}},\ \bibinfo {pages} {10267} (\bibinfo {year}
  {2000})}\BibitemShut {NoStop}%
\bibitem [{\citenamefont {Kitaev}(2001)}]{Kitaev2001}%
  \BibitemOpen
  \bibfield  {author} {\bibinfo {author} {\bibfnamefont {A.}~\bibnamefont
  {Kitaev}},\ }\href@noop {} {\bibfield  {journal} {\bibinfo  {journal} {Phys.
  Usp.}\ }\textbf {\bibinfo {volume} {44}},\ \bibinfo {pages} {131} (\bibinfo
  {year} {2001})}\BibitemShut {NoStop}%
\bibitem [{\citenamefont {Saxena}\ \emph {et~al.}(2000)\citenamefont {Saxena},
  \citenamefont {Agarwal}, \citenamefont {Ahilan}, \citenamefont {Grosche},
  \citenamefont {Haselwimmer}, \citenamefont {Steiner}, \citenamefont {Pugh},
  \citenamefont {Walker}, \citenamefont {Julian}, \citenamefont {Monthoux},
  \citenamefont {Lonzarich}, \citenamefont {Huxley}, \citenamefont {Sheikin},
  \citenamefont {Braithwaite},\ and\ \citenamefont {Flouquet}}]{Saxena2000}%
  \BibitemOpen
  \bibfield  {author} {\bibinfo {author} {\bibfnamefont {S.~S.}\ \bibnamefont
  {Saxena}}, \bibinfo {author} {\bibfnamefont {P.}~\bibnamefont {Agarwal}},
  \bibinfo {author} {\bibfnamefont {K.}~\bibnamefont {Ahilan}}, \bibinfo
  {author} {\bibfnamefont {F.~M.}\ \bibnamefont {Grosche}}, \bibinfo {author}
  {\bibfnamefont {R.~K.~W.}\ \bibnamefont {Haselwimmer}}, \bibinfo {author}
  {\bibfnamefont {M.~J.}\ \bibnamefont {Steiner}}, \bibinfo {author}
  {\bibfnamefont {E.}~\bibnamefont {Pugh}}, \bibinfo {author} {\bibfnamefont
  {I.~R.}\ \bibnamefont {Walker}}, \bibinfo {author} {\bibfnamefont {S.~R.}\
  \bibnamefont {Julian}}, \bibinfo {author} {\bibfnamefont {P.}~\bibnamefont
  {Monthoux}}, \bibinfo {author} {\bibfnamefont {G.~G.}\ \bibnamefont
  {Lonzarich}}, \bibinfo {author} {\bibfnamefont {A.}~\bibnamefont {Huxley}},
  \bibinfo {author} {\bibfnamefont {I.}~\bibnamefont {Sheikin}}, \bibinfo
  {author} {\bibfnamefont {D.}~\bibnamefont {Braithwaite}}, \ and\ \bibinfo
  {author} {\bibfnamefont {J.}~\bibnamefont {Flouquet}},\ }\href@noop {}
  {\bibfield  {journal} {\bibinfo  {journal} {Nature}\ }\textbf {\bibinfo
  {volume} {406}},\ \bibinfo {pages} {587} (\bibinfo {year}
  {2000})}\BibitemShut {NoStop}%
\bibitem [{\citenamefont {Aoki}\ \emph {et~al.}(2001)\citenamefont {Aoki},
  \citenamefont {Huxley}, \citenamefont {Ressouche}, \citenamefont
  {Braithwaite}, \citenamefont {Flouquet}, \citenamefont {Brison},
  \citenamefont {Lhotel},\ and\ \citenamefont {Paulsen}}]{Aoki2001}%
  \BibitemOpen
  \bibfield  {author} {\bibinfo {author} {\bibfnamefont {D.}~\bibnamefont
  {Aoki}}, \bibinfo {author} {\bibfnamefont {A.}~\bibnamefont {Huxley}},
  \bibinfo {author} {\bibfnamefont {E.}~\bibnamefont {Ressouche}}, \bibinfo
  {author} {\bibfnamefont {D.}~\bibnamefont {Braithwaite}}, \bibinfo {author}
  {\bibfnamefont {J.}~\bibnamefont {Flouquet}}, \bibinfo {author}
  {\bibfnamefont {J.-P.}\ \bibnamefont {Brison}}, \bibinfo {author}
  {\bibfnamefont {E.}~\bibnamefont {Lhotel}}, \ and\ \bibinfo {author}
  {\bibfnamefont {C.}~\bibnamefont {Paulsen}},\ }\href@noop {} {\bibfield
  {journal} {\bibinfo  {journal} {Nature}\ }\textbf {\bibinfo {volume} {413}},\
  \bibinfo {pages} {613} (\bibinfo {year} {2001})}\BibitemShut {NoStop}%
\bibitem [{\citenamefont {Huy}\ \emph {et~al.}(2007)\citenamefont {Huy},
  \citenamefont {Gasparini}, \citenamefont {de~Nijs}, \citenamefont {Huang},
  \citenamefont {Klaasse}, \citenamefont {Gortenmulder}, \citenamefont
  {de~Visser}, \citenamefont {Hamann}, \citenamefont {Gorlach},\ and\
  \citenamefont {v.~Lohneysen}}]{Huy2007}%
  \BibitemOpen
  \bibfield  {author} {\bibinfo {author} {\bibfnamefont {N.~T.}\ \bibnamefont
  {Huy}}, \bibinfo {author} {\bibfnamefont {A.}~\bibnamefont {Gasparini}},
  \bibinfo {author} {\bibfnamefont {D.~E.}\ \bibnamefont {de~Nijs}}, \bibinfo
  {author} {\bibfnamefont {Y.}~\bibnamefont {Huang}}, \bibinfo {author}
  {\bibfnamefont {J.~C.~P.}\ \bibnamefont {Klaasse}}, \bibinfo {author}
  {\bibfnamefont {T.}~\bibnamefont {Gortenmulder}}, \bibinfo {author}
  {\bibfnamefont {A.}~\bibnamefont {de~Visser}}, \bibinfo {author}
  {\bibfnamefont {A.}~\bibnamefont {Hamann}}, \bibinfo {author} {\bibfnamefont
  {T.}~\bibnamefont {Gorlach}}, \ and\ \bibinfo {author} {\bibfnamefont
  {H.}~\bibnamefont {v.~Lohneysen}},\ }\href@noop {} {\bibfield  {journal}
  {\bibinfo  {journal} {Phys. Rev. Lett.}\ }\textbf {\bibinfo {volume} {99}},\
  \bibinfo {pages} {067006} (\bibinfo {year} {2007})}\BibitemShut {NoStop}%
\bibitem [{\citenamefont {Bergeret}\ \emph {et~al.}(2001)\citenamefont
  {Bergeret}, \citenamefont {Volkov},\ and\ \citenamefont
  {Efetov}}]{Efetov2001}%
  \BibitemOpen
  \bibfield  {author} {\bibinfo {author} {\bibfnamefont {F.~S.}\ \bibnamefont
  {Bergeret}}, \bibinfo {author} {\bibfnamefont {A.~F.}\ \bibnamefont
  {Volkov}}, \ and\ \bibinfo {author} {\bibfnamefont {K.~B.}\ \bibnamefont
  {Efetov}},\ }\href@noop {} {\bibfield  {journal} {\bibinfo  {journal} {Phys.
  Rev. Lett.}\ }\textbf {\bibinfo {volume} {86}},\ \bibinfo {pages} {4096}
  (\bibinfo {year} {2001})}\BibitemShut {NoStop}%
\bibitem [{\citenamefont {Kadigrobov}\ \emph {et~al.}(2001)\citenamefont
  {Kadigrobov}, \citenamefont {Shekhter},\ and\ \citenamefont
  {Jonson}}]{Jonson2001}%
  \BibitemOpen
  \bibfield  {author} {\bibinfo {author} {\bibfnamefont {A.}~\bibnamefont
  {Kadigrobov}}, \bibinfo {author} {\bibfnamefont {R.~I.}\ \bibnamefont
  {Shekhter}}, \ and\ \bibinfo {author} {\bibfnamefont {M.}~\bibnamefont
  {Jonson}},\ }\href@noop {} {\bibfield  {journal} {\bibinfo  {journal}
  {Europhysics Letters}\ }\textbf {\bibinfo {volume} {54 (3)}},\ \bibinfo
  {pages} {394} (\bibinfo {year} {2001})}\BibitemShut {NoStop}%
\bibitem [{\citenamefont {Law}\ \emph {et~al.}(2009)\citenamefont {Law},
  \citenamefont {Lee},\ and\ \citenamefont {Ng}}]{Ng2009}%
  \BibitemOpen
  \bibfield  {author} {\bibinfo {author} {\bibfnamefont {K.~T.}\ \bibnamefont
  {Law}}, \bibinfo {author} {\bibfnamefont {P.~A.}\ \bibnamefont {Lee}}, \ and\
  \bibinfo {author} {\bibfnamefont {T.~K.}\ \bibnamefont {Ng}},\ }\href@noop {}
  {\bibfield  {journal} {\bibinfo  {journal} {Phys. Rev. Lett.}\ }\textbf
  {\bibinfo {volume} {103}},\ \bibinfo {pages} {237001} (\bibinfo {year}
  {2009})}\BibitemShut {NoStop}%
\bibitem [{\citenamefont {Gong}\ \emph {et~al.}(2015)\citenamefont {Gong},
  \citenamefont {He-Xin}, \citenamefont {Xu}, \citenamefont {Yue},
  \citenamefont {Zhu}, \citenamefont {Jin}, \citenamefont {Tian},\ and\
  \citenamefont {Zhao}}]{Jin2015}%
  \BibitemOpen
  \bibfield  {author} {\bibinfo {author} {\bibfnamefont {X.-X.}\ \bibnamefont
  {Gong}}, \bibinfo {author} {\bibnamefont {He-Xin}}, \bibinfo {author}
  {\bibfnamefont {Z.~P.-C.}\ \bibnamefont {Xu}}, \bibinfo {author}
  {\bibfnamefont {D.}~\bibnamefont {Yue}}, \bibinfo {author} {\bibfnamefont
  {K.}~\bibnamefont {Zhu}}, \bibinfo {author} {\bibfnamefont {X.-F.}\
  \bibnamefont {Jin}}, \bibinfo {author} {\bibfnamefont {H.}~\bibnamefont
  {Tian}}, \ and\ \bibinfo {author} {\bibfnamefont {G.-J.}\ \bibnamefont
  {Zhao}},\ }\href@noop {} {\bibfield  {journal} {\bibinfo  {journal} {Chinese
  Physics Letter}\ }\textbf {\bibinfo {volume} {32}},\ \bibinfo {pages}
  {067402} (\bibinfo {year} {2015})}\BibitemShut {NoStop}%
\bibitem [{\citenamefont {Moodera}\ and\ \citenamefont
  {Meservey}(1990)}]{Moodera1990}%
  \BibitemOpen
  \bibfield  {author} {\bibinfo {author} {\bibfnamefont {J.~S.}\ \bibnamefont
  {Moodera}}\ and\ \bibinfo {author} {\bibfnamefont {R.}~\bibnamefont
  {Meservey}},\ }\href@noop {} {\bibfield  {journal} {\bibinfo  {journal}
  {Phys. Rev. B}\ }\textbf {\bibinfo {volume} {42}},\ \bibinfo {pages} {179}
  (\bibinfo {year} {1990})}\BibitemShut {NoStop}%
\bibitem [{\citenamefont {LeClair}\ \emph {et~al.}(2005)\citenamefont
  {LeClair}, \citenamefont {Moodera}, \citenamefont {Philip},\ and\
  \citenamefont {Heiman}}]{Heiman2005}%
  \BibitemOpen
  \bibfield  {author} {\bibinfo {author} {\bibfnamefont {P.}~\bibnamefont
  {LeClair}}, \bibinfo {author} {\bibfnamefont {J.~S.}\ \bibnamefont
  {Moodera}}, \bibinfo {author} {\bibfnamefont {J.}~\bibnamefont {Philip}}, \
  and\ \bibinfo {author} {\bibfnamefont {D.}~\bibnamefont {Heiman}},\
  }\href@noop {} {\bibfield  {journal} {\bibinfo  {journal} {Phys. Rev. Lett}\
  }\textbf {\bibinfo {volume} {94}},\ \bibinfo {pages} {037006} (\bibinfo
  {year} {2005})}\BibitemShut {NoStop}%
\bibitem [{\citenamefont {Prakash}\ \emph {et~al.}(2017)\citenamefont
  {Prakash}, \citenamefont {Kumar}, \citenamefont {Thamizhavel},\ and\
  \citenamefont {Ramakrishnan}}]{Prakash2017}%
  \BibitemOpen
  \bibfield  {author} {\bibinfo {author} {\bibfnamefont {O.}~\bibnamefont
  {Prakash}}, \bibinfo {author} {\bibfnamefont {A.}~\bibnamefont {Kumar}},
  \bibinfo {author} {\bibfnamefont {A.}~\bibnamefont {Thamizhavel}}, \ and\
  \bibinfo {author} {\bibfnamefont {S.}~\bibnamefont {Ramakrishnan}},\
  }\href@noop {} {\bibfield  {journal} {\bibinfo  {journal} {Science}\ }\textbf
  {\bibinfo {volume} {355}},\ \bibinfo {pages} {52} (\bibinfo {year}
  {2017})}\BibitemShut {NoStop}%
\bibitem [{\citenamefont {Siva}\ \emph {et~al.}(2015)\citenamefont {Siva},
  \citenamefont {Senapati}, \citenamefont {Satpati}, \citenamefont {Prusty},
  \citenamefont {Avasthi}, \citenamefont {Kanjilal},\ and\ \citenamefont
  {Sahoo}}]{Pratap2015}%
  \BibitemOpen
  \bibfield  {author} {\bibinfo {author} {\bibfnamefont {V.}~\bibnamefont
  {Siva}}, \bibinfo {author} {\bibfnamefont {K.}~\bibnamefont {Senapati}},
  \bibinfo {author} {\bibfnamefont {B.}~\bibnamefont {Satpati}}, \bibinfo
  {author} {\bibfnamefont {S.}~\bibnamefont {Prusty}}, \bibinfo {author}
  {\bibfnamefont {D.~K.}\ \bibnamefont {Avasthi}}, \bibinfo {author}
  {\bibfnamefont {D.}~\bibnamefont {Kanjilal}}, \ and\ \bibinfo {author}
  {\bibfnamefont {P.~K.}\ \bibnamefont {Sahoo}},\ }\href@noop {} {\bibfield
  {journal} {\bibinfo  {journal} {Journal of Applied Physics}\ }\textbf
  {\bibinfo {volume} {117}},\ \bibinfo {pages} {083902} (\bibinfo {year}
  {2015})}\BibitemShut {NoStop}%
\bibitem [{\citenamefont {Chao}\ \emph {et~al.}(2017)\citenamefont {Chao},
  \citenamefont {Lu}, \citenamefont {Lin}, \citenamefont {Chiu}, \citenamefont
  {Li}, \citenamefont {Chen}, \citenamefont {Liou},\ and\ \citenamefont
  {Lee}}]{Chao2017}%
  \BibitemOpen
  \bibfield  {author} {\bibinfo {author} {\bibfnamefont {S.}~\bibnamefont
  {Chao}}, \bibinfo {author} {\bibfnamefont {S.~C.}\ \bibnamefont {Lu}},
  \bibinfo {author} {\bibfnamefont {J.~H.}\ \bibnamefont {Lin}}, \bibinfo
  {author} {\bibfnamefont {P.}~\bibnamefont {Chiu}}, \bibinfo {author}
  {\bibfnamefont {W.~J.}\ \bibnamefont {Li}}, \bibinfo {author} {\bibfnamefont
  {P.~J.}\ \bibnamefont {Chen}}, \bibinfo {author} {\bibfnamefont
  {Y.}~\bibnamefont {Liou}}, \ and\ \bibinfo {author} {\bibfnamefont {T.~K.}\
  \bibnamefont {Lee}},\ }\href@noop {} {\bibfield  {journal} {\bibinfo
  {journal} {APS March Meeting 2017}\ }\textbf {\bibinfo {volume} {62}},\
  \bibinfo {pages} {R45.00006} (\bibinfo {year} {2017})}\BibitemShut {NoStop}%
\bibitem [{\citenamefont {Liu}\ \emph {et~al.}(2018)\citenamefont {Liu},
  \citenamefont {Xing}, \citenamefont {Merino}, \citenamefont {Micklitz},
  \citenamefont {Franceschini}, \citenamefont {Baggio-Saitovitch},
  \citenamefont {Bell},\ and\ \citenamefont {Solorzano}}]{Liu2018}%
  \BibitemOpen
  \bibfield  {author} {\bibinfo {author} {\bibfnamefont {L.~Y.}\ \bibnamefont
  {Liu}}, \bibinfo {author} {\bibfnamefont {Y.~T.}\ \bibnamefont {Xing}},
  \bibinfo {author} {\bibfnamefont {I.~L.~C.}\ \bibnamefont {Merino}}, \bibinfo
  {author} {\bibfnamefont {H.}~\bibnamefont {Micklitz}}, \bibinfo {author}
  {\bibfnamefont {D.~F.}\ \bibnamefont {Franceschini}}, \bibinfo {author}
  {\bibfnamefont {E.}~\bibnamefont {Baggio-Saitovitch}}, \bibinfo {author}
  {\bibfnamefont {D.~C.}\ \bibnamefont {Bell}}, \ and\ \bibinfo {author}
  {\bibfnamefont {I.~G.}\ \bibnamefont {Solorzano}},\ }\href@noop {} {\bibfield
   {journal} {\bibinfo  {journal} {Phys. Rev. Materials.}\ }\textbf {\bibinfo
  {volume} {2}},\ \bibinfo {pages} {014601} (\bibinfo {year}
  {2018})}\BibitemShut {NoStop}%
\bibitem [{\citenamefont {Kumar}\ \emph {et~al.}(2011)\citenamefont {Kumar},
  \citenamefont {Kumar}, \citenamefont {Vajpayee}, \citenamefont {Gahtori},
  \citenamefont {Sharma}, \citenamefont {Ahluwalia}, \citenamefont {Auluck},\
  and\ \citenamefont {Awana}}]{Awana2011}%
  \BibitemOpen
  \bibfield  {author} {\bibinfo {author} {\bibfnamefont {J.}~\bibnamefont
  {Kumar}}, \bibinfo {author} {\bibfnamefont {A.}~\bibnamefont {Kumar}},
  \bibinfo {author} {\bibfnamefont {A.}~\bibnamefont {Vajpayee}}, \bibinfo
  {author} {\bibfnamefont {B.}~\bibnamefont {Gahtori}}, \bibinfo {author}
  {\bibfnamefont {D.}~\bibnamefont {Sharma}}, \bibinfo {author} {\bibfnamefont
  {P.~K.}\ \bibnamefont {Ahluwalia}}, \bibinfo {author} {\bibfnamefont
  {S.}~\bibnamefont {Auluck}}, \ and\ \bibinfo {author} {\bibfnamefont
  {V.~P.~S.}\ \bibnamefont {Awana}},\ }\href@noop {} {\bibfield  {journal}
  {\bibinfo  {journal} {Supercond. Sci. Technol.}\ }\textbf {\bibinfo {volume}
  {24}},\ \bibinfo {pages} {085002} (\bibinfo {year} {2011})}\BibitemShut
  {NoStop}%
\bibitem [{\citenamefont {Sau}\ \emph {et~al.}(2010)\citenamefont {Sau},
  \citenamefont {Lutchyn}, \citenamefont {Tewari},\ and\ \citenamefont
  {DasSarma}}]{Sau2010}%
  \BibitemOpen
  \bibfield  {author} {\bibinfo {author} {\bibfnamefont {J.~D.}\ \bibnamefont
  {Sau}}, \bibinfo {author} {\bibfnamefont {R.~M.}\ \bibnamefont {Lutchyn}},
  \bibinfo {author} {\bibfnamefont {S.}~\bibnamefont {Tewari}}, \ and\ \bibinfo
  {author} {\bibfnamefont {S.}~\bibnamefont {DasSarma}},\ }\href@noop {}
  {\bibfield  {journal} {\bibinfo  {journal} {Phys. Rev. Lett.}\ }\textbf
  {\bibinfo {volume} {104}},\ \bibinfo {pages} {040502} (\bibinfo {year}
  {2010})}\BibitemShut {NoStop}%
\bibitem [{\citenamefont {Alicea}(2010)}]{Jason2010}%
  \BibitemOpen
  \bibfield  {author} {\bibinfo {author} {\bibfnamefont {J.}~\bibnamefont
  {Alicea}},\ }\href@noop {} {\bibfield  {journal} {\bibinfo  {journal} {Phys.
  Rev.B}\ }\textbf {\bibinfo {volume} {81}},\ \bibinfo {pages} {125318}
  (\bibinfo {year} {2010})}\BibitemShut {NoStop}%
\bibitem [{\citenamefont {Fu}\ and\ \citenamefont {Kane}(2008)}]{Kane2008}%
  \BibitemOpen
  \bibfield  {author} {\bibinfo {author} {\bibfnamefont {L.}~\bibnamefont
  {Fu}}\ and\ \bibinfo {author} {\bibfnamefont {C.~L.}\ \bibnamefont {Kane}},\
  }\href@noop {} {\bibfield  {journal} {\bibinfo  {journal} {Phys. Rev. Lett.}\
  }\textbf {\bibinfo {volume} {100}},\ \bibinfo {pages} {096407} (\bibinfo
  {year} {2008})}\BibitemShut {NoStop}%
\bibitem [{\citenamefont {van Hulst}\ \emph {et~al.}(1993)\citenamefont {van
  Hulst}, \citenamefont {Rietveld}, \citenamefont {van~der Marel},
  \citenamefont {Tuinstra},\ and\ \citenamefont {Jaeger}}]{Jaeger93}%
  \BibitemOpen
  \bibfield  {author} {\bibinfo {author} {\bibfnamefont {J.~A.}\ \bibnamefont
  {van Hulst}}, \bibinfo {author} {\bibfnamefont {G.}~\bibnamefont {Rietveld}},
  \bibinfo {author} {\bibfnamefont {D.}~\bibnamefont {van~der Marel}}, \bibinfo
  {author} {\bibfnamefont {F.}~\bibnamefont {Tuinstra}}, \ and\ \bibinfo
  {author} {\bibfnamefont {H.~M.}\ \bibnamefont {Jaeger}},\ }\href@noop {}
  {\bibfield  {journal} {\bibinfo  {journal} {Phys. Rev. B}\ }\textbf {\bibinfo
  {volume} {47}},\ \bibinfo {pages} {548} (\bibinfo {year} {1993})}\BibitemShut
  {NoStop}%
\bibitem [{\citenamefont {Zhou}\ \emph {et~al.}(2017)\citenamefont {Zhou},
  \citenamefont {Gong},\ and\ \citenamefont {Jin}}]{Zhou2017}%
  \BibitemOpen
  \bibfield  {author} {\bibinfo {author} {\bibfnamefont {H.}~\bibnamefont
  {Zhou}}, \bibinfo {author} {\bibfnamefont {X.}~\bibnamefont {Gong}}, \ and\
  \bibinfo {author} {\bibfnamefont {X.}~\bibnamefont {Jin}},\ }\href@noop {}
  {\bibfield  {journal} {\bibinfo  {journal} {Journal of Magnetism and Magnetic
  Materials}\ }\textbf {\bibinfo {volume} {422}},\ \bibinfo {pages} {73}
  (\bibinfo {year} {2017})}\BibitemShut {NoStop}%
\bibitem [{\citenamefont {Zhao}\ \emph {et~al.}(2018)\citenamefont {Zhao},
  \citenamefont {Gong}, \citenamefont {Xu}, \citenamefont {Li}, \citenamefont
  {Huang}, \citenamefont {Jin}, \citenamefont {Zhu},\ and\ \citenamefont
  {Chen}}]{Zhao2018}%
  \BibitemOpen
  \bibfield  {author} {\bibinfo {author} {\bibfnamefont {G.~J.}\ \bibnamefont
  {Zhao}}, \bibinfo {author} {\bibfnamefont {X.~X.}\ \bibnamefont {Gong}},
  \bibinfo {author} {\bibfnamefont {P.~C.}\ \bibnamefont {Xu}}, \bibinfo
  {author} {\bibfnamefont {B.~C.}\ \bibnamefont {Li}}, \bibinfo {author}
  {\bibfnamefont {Z.~Y.}\ \bibnamefont {Huang}}, \bibinfo {author}
  {\bibfnamefont {X.~F.}\ \bibnamefont {Jin}}, \bibinfo {author} {\bibfnamefont
  {X.~D.}\ \bibnamefont {Zhu}}, \ and\ \bibinfo {author} {\bibfnamefont
  {T.~Y.}\ \bibnamefont {Chen}},\ }\href@noop {} {\bibfield  {journal}
  {\bibinfo  {journal} {arXiv: 1805.09811}\ } (\bibinfo {year}
  {2018})}\BibitemShut {NoStop}%
\bibitem [{\citenamefont {Chauhan}\ \emph {et~al.}(2019)\citenamefont
  {Chauhan}, \citenamefont {Mahmood}, \citenamefont {Yue}, \citenamefont {Xu},
  \citenamefont {Jin},\ and\ \citenamefont {Armitage}}]{Armitage2019}%
  \BibitemOpen
  \bibfield  {author} {\bibinfo {author} {\bibfnamefont {P.}~\bibnamefont
  {Chauhan}}, \bibinfo {author} {\bibfnamefont {F.}~\bibnamefont {Mahmood}},
  \bibinfo {author} {\bibfnamefont {D.}~\bibnamefont {Yue}}, \bibinfo {author}
  {\bibfnamefont {P.-C.}\ \bibnamefont {Xu}}, \bibinfo {author} {\bibfnamefont
  {X.}~\bibnamefont {Jin}}, \ and\ \bibinfo {author} {\bibfnamefont
  {N.}~\bibnamefont {Armitage}},\ }\href@noop {} {\bibfield  {journal}
  {\bibinfo  {journal} {Phys. Rev. Lett.}\ }\textbf {\bibinfo {volume} {122}},\
  \bibinfo {pages} {017002} (\bibinfo {year} {2019})}\BibitemShut {NoStop}%
\bibitem [{\citenamefont {Xiao}\ \emph {et~al.}(2012)\citenamefont {Xiao},
  \citenamefont {Wei},\ and\ \citenamefont {Jin}}]{Jin2012}%
  \BibitemOpen
  \bibfield  {author} {\bibinfo {author} {\bibfnamefont {S.}~\bibnamefont
  {Xiao}}, \bibinfo {author} {\bibfnamefont {D.}~\bibnamefont {Wei}}, \ and\
  \bibinfo {author} {\bibfnamefont {X.}~\bibnamefont {Jin}},\ }\href@noop {}
  {\bibfield  {journal} {\bibinfo  {journal} {Phys. Rev. Lett.}\ }\textbf
  {\bibinfo {volume} {109}},\ \bibinfo {pages} {166805} (\bibinfo {year}
  {2012})}\BibitemShut {NoStop}%
\bibitem [{\citenamefont {Koroteev}\ \emph {et~al.}(2004)\citenamefont
  {Koroteev}, \citenamefont {Bihlmayer}, \citenamefont {Gayone}, \citenamefont
  {Chulkov}, \citenamefont {Blugel}, \citenamefont {Echenique},\ and\
  \citenamefont {Hofmann}}]{Hofmann2004}%
  \BibitemOpen
  \bibfield  {author} {\bibinfo {author} {\bibfnamefont {Y.~M.}\ \bibnamefont
  {Koroteev}}, \bibinfo {author} {\bibfnamefont {G.}~\bibnamefont {Bihlmayer}},
  \bibinfo {author} {\bibfnamefont {J.~E.}\ \bibnamefont {Gayone}}, \bibinfo
  {author} {\bibfnamefont {E.~V.}\ \bibnamefont {Chulkov}}, \bibinfo {author}
  {\bibfnamefont {S.}~\bibnamefont {Blugel}}, \bibinfo {author} {\bibfnamefont
  {P.~M.}\ \bibnamefont {Echenique}}, \ and\ \bibinfo {author} {\bibfnamefont
  {P.}~\bibnamefont {Hofmann}},\ }\href@noop {} {\bibfield  {journal} {\bibinfo
   {journal} {Phys. Rev. Lett.}\ }\textbf {\bibinfo {volume} {93}},\ \bibinfo
  {pages} {046403} (\bibinfo {year} {2004})}\BibitemShut {NoStop}%
\bibitem [{\citenamefont {Hirahara}\ \emph {et~al.}(2006)\citenamefont
  {Hirahara}, \citenamefont {Nagao}, \citenamefont {Matsuda}, \citenamefont
  {Bihlmayer}, \citenamefont {Chulkov}, \citenamefont {Koroteev}, \citenamefont
  {Echenique}, \citenamefont {Saito},\ and\ \citenamefont
  {Hasegawa}}]{Hasegawa2006}%
  \BibitemOpen
  \bibfield  {author} {\bibinfo {author} {\bibfnamefont {T.}~\bibnamefont
  {Hirahara}}, \bibinfo {author} {\bibfnamefont {T.}~\bibnamefont {Nagao}},
  \bibinfo {author} {\bibfnamefont {I.}~\bibnamefont {Matsuda}}, \bibinfo
  {author} {\bibfnamefont {G.}~\bibnamefont {Bihlmayer}}, \bibinfo {author}
  {\bibfnamefont {E.~V.}\ \bibnamefont {Chulkov}}, \bibinfo {author}
  {\bibfnamefont {Y.~M.}\ \bibnamefont {Koroteev}}, \bibinfo {author}
  {\bibfnamefont {P.~M.}\ \bibnamefont {Echenique}}, \bibinfo {author}
  {\bibfnamefont {M.}~\bibnamefont {Saito}}, \ and\ \bibinfo {author}
  {\bibfnamefont {S.}~\bibnamefont {Hasegawa}},\ }\href@noop {} {\bibfield
  {journal} {\bibinfo  {journal} {Phys. Rev. Lett.}\ }\textbf {\bibinfo
  {volume} {97}},\ \bibinfo {pages} {146803} (\bibinfo {year}
  {2006})}\BibitemShut {NoStop}%
\bibitem [{\citenamefont {Saito}\ \emph {et~al.}(2016)\citenamefont {Saito},
  \citenamefont {Sawahata}, \citenamefont {Komine},\ and\ \citenamefont
  {Aono}}]{Aono2016}%
  \BibitemOpen
  \bibfield  {author} {\bibinfo {author} {\bibfnamefont {K.}~\bibnamefont
  {Saito}}, \bibinfo {author} {\bibfnamefont {H.}~\bibnamefont {Sawahata}},
  \bibinfo {author} {\bibfnamefont {T.}~\bibnamefont {Komine}}, \ and\ \bibinfo
  {author} {\bibfnamefont {T.}~\bibnamefont {Aono}},\ }\href@noop {} {\bibfield
   {journal} {\bibinfo  {journal} {Phys. Rev. B}\ }\textbf {\bibinfo {volume}
  {93}},\ \bibinfo {pages} {041301(R)} (\bibinfo {year} {2016})}\BibitemShut
  {NoStop}%
\bibitem [{\citenamefont {Koroteev}\ \emph {et~al.}(2008)\citenamefont
  {Koroteev}, \citenamefont {Bihlmayer}, \citenamefont {Chulkov},\ and\
  \citenamefont {Blugel}}]{Koroteev2008}%
  \BibitemOpen
  \bibfield  {author} {\bibinfo {author} {\bibfnamefont {Y.~M.}\ \bibnamefont
  {Koroteev}}, \bibinfo {author} {\bibfnamefont {G.}~\bibnamefont {Bihlmayer}},
  \bibinfo {author} {\bibfnamefont {E.~V.}\ \bibnamefont {Chulkov}}, \ and\
  \bibinfo {author} {\bibfnamefont {S.}~\bibnamefont {Blugel}},\ }\href@noop {}
  {\bibfield  {journal} {\bibinfo  {journal} {Phys. Rev. B.}\ }\textbf
  {\bibinfo {volume} {77}},\ \bibinfo {pages} {045428} (\bibinfo {year}
  {2008})}\BibitemShut {NoStop}%
\bibitem [{\citenamefont {Hirahara}\ \emph {et~al.}(2007)\citenamefont
  {Hirahara}, \citenamefont {Miyamoto}, \citenamefont {Matsuda}, \citenamefont
  {Kadono}, \citenamefont {Kimura}, \citenamefont {Nagao}, \citenamefont
  {Bihlmayer}, \citenamefont {Chulkov}, \citenamefont {Qiao}, \citenamefont
  {Shimada}, \citenamefont {Namatame}, \citenamefont {Taniguchi},\ and\
  \citenamefont {Hasegawa}}]{Hirahara2007}%
  \BibitemOpen
  \bibfield  {author} {\bibinfo {author} {\bibfnamefont {T.}~\bibnamefont
  {Hirahara}}, \bibinfo {author} {\bibfnamefont {K.}~\bibnamefont {Miyamoto}},
  \bibinfo {author} {\bibfnamefont {I.}~\bibnamefont {Matsuda}}, \bibinfo
  {author} {\bibfnamefont {T.}~\bibnamefont {Kadono}}, \bibinfo {author}
  {\bibfnamefont {A.}~\bibnamefont {Kimura}}, \bibinfo {author} {\bibfnamefont
  {T.}~\bibnamefont {Nagao}}, \bibinfo {author} {\bibfnamefont
  {G.}~\bibnamefont {Bihlmayer}}, \bibinfo {author} {\bibfnamefont {E.~V.}\
  \bibnamefont {Chulkov}}, \bibinfo {author} {\bibfnamefont {S.}~\bibnamefont
  {Qiao}}, \bibinfo {author} {\bibfnamefont {K.}~\bibnamefont {Shimada}},
  \bibinfo {author} {\bibfnamefont {H.}~\bibnamefont {Namatame}}, \bibinfo
  {author} {\bibfnamefont {M.}~\bibnamefont {Taniguchi}}, \ and\ \bibinfo
  {author} {\bibfnamefont {S.}~\bibnamefont {Hasegawa}},\ }\href@noop {}
  {\bibfield  {journal} {\bibinfo  {journal} {Phys. Rev. B.}\ }\textbf
  {\bibinfo {volume} {76}},\ \bibinfo {pages} {153305} (\bibinfo {year}
  {2007})}\BibitemShut {NoStop}%
\bibitem [{\citenamefont {Pascual}\ \emph {et~al.}(2004)\citenamefont
  {Pascual}, \citenamefont {Bihlmayer}, \citenamefont {Koroteev}, \citenamefont
  {Rust}, \citenamefont {Ceballos}, \citenamefont {Hansmann}, \citenamefont
  {Horn}, \citenamefont {Chulkov}, \citenamefont {Blugel}, \citenamefont
  {Echenique},\ and\ \citenamefont {Hofmann}}]{Pascual2004}%
  \BibitemOpen
  \bibfield  {author} {\bibinfo {author} {\bibfnamefont {J.~I.}\ \bibnamefont
  {Pascual}}, \bibinfo {author} {\bibfnamefont {G.}~\bibnamefont {Bihlmayer}},
  \bibinfo {author} {\bibfnamefont {Y.~M.}\ \bibnamefont {Koroteev}}, \bibinfo
  {author} {\bibfnamefont {H.-P.}\ \bibnamefont {Rust}}, \bibinfo {author}
  {\bibfnamefont {G.}~\bibnamefont {Ceballos}}, \bibinfo {author}
  {\bibfnamefont {M.}~\bibnamefont {Hansmann}}, \bibinfo {author}
  {\bibfnamefont {K.}~\bibnamefont {Horn}}, \bibinfo {author} {\bibfnamefont
  {E.~V.}\ \bibnamefont {Chulkov}}, \bibinfo {author} {\bibfnamefont
  {S.}~\bibnamefont {Blugel}}, \bibinfo {author} {\bibfnamefont {P.~M.}\
  \bibnamefont {Echenique}}, \ and\ \bibinfo {author} {\bibfnamefont
  {P.}~\bibnamefont {Hofmann}},\ }\href@noop {} {\bibfield  {journal} {\bibinfo
   {journal} {Phys. Rev. Lett.}\ }\textbf {\bibinfo {volume} {93}},\ \bibinfo
  {pages} {196802} (\bibinfo {year} {2004})}\BibitemShut {NoStop}%
\bibitem [{\citenamefont {Yokoyama}\ and\ \citenamefont
  {Tanaka}(2007)}]{Tanaka2007}%
  \BibitemOpen
  \bibfield  {author} {\bibinfo {author} {\bibfnamefont {T.}~\bibnamefont
  {Yokoyama}}\ and\ \bibinfo {author} {\bibfnamefont {Y.}~\bibnamefont
  {Tanaka}},\ }\href@noop {} {\bibfield  {journal} {\bibinfo  {journal} {Phys.
  Rev. B}\ }\textbf {\bibinfo {volume} {75}},\ \bibinfo {pages} {132503}
  (\bibinfo {year} {2007})}\BibitemShut {NoStop}%
\bibitem [{\citenamefont {Jin}(2016)}]{Jintalk2016}%
  \BibitemOpen
  \bibfield  {author} {\bibinfo {author} {\bibfnamefont {X.}~\bibnamefont
  {Jin}},\ }\href@noop {} {\bibfield  {journal} {\bibinfo  {journal} {talk
  given in a seminar at Institute of Physics, Academia Sinica, Taiwan.}\ }
  (\bibinfo {year} {2016})}\BibitemShut {NoStop}%
\bibitem [{\citenamefont {Chien}(2018)}]{Chientalk2018}%
  \BibitemOpen
  \bibfield  {author} {\bibinfo {author} {\bibfnamefont {C.~L.}\ \bibnamefont
  {Chien}},\ }\href@noop {} {\bibfield  {journal} {\bibinfo  {journal} {talk
  titled "Triplet p-wave superconductivity with ABM state in Bi/Ni bilayers"
  given in the Institute of Physics, Academia Sinica, Taiwan.}\ } (\bibinfo
  {year} {2018})}\BibitemShut {NoStop}%
\bibitem [{\citenamefont {Gong}\ \emph {et~al.}(2017)\citenamefont {Gong},
  \citenamefont {Kargarian}, \citenamefont {Stern}, \citenamefont {Yue},
  \citenamefont {Zhou}, \citenamefont {Jin}, \citenamefont {Galitski},
  \citenamefont {Yakovenko},\ and\ \citenamefont {Xia}}]{Xia2017}%
  \BibitemOpen
  \bibfield  {author} {\bibinfo {author} {\bibfnamefont {X.}~\bibnamefont
  {Gong}}, \bibinfo {author} {\bibfnamefont {M.}~\bibnamefont {Kargarian}},
  \bibinfo {author} {\bibfnamefont {A.}~\bibnamefont {Stern}}, \bibinfo
  {author} {\bibfnamefont {D.}~\bibnamefont {Yue}}, \bibinfo {author}
  {\bibfnamefont {H.}~\bibnamefont {Zhou}}, \bibinfo {author} {\bibfnamefont
  {X.}~\bibnamefont {Jin}}, \bibinfo {author} {\bibfnamefont {V.~M.}\
  \bibnamefont {Galitski}}, \bibinfo {author} {\bibfnamefont {V.~M.}\
  \bibnamefont {Yakovenko}}, \ and\ \bibinfo {author} {\bibfnamefont
  {J.}~\bibnamefont {Xia}},\ }\href@noop {} {\bibfield  {journal} {\bibinfo
  {journal} {Science Advances}\ }\textbf {\bibinfo {volume} {3}},\ \bibinfo
  {pages} {e1602579} (\bibinfo {year} {2017})}\BibitemShut {NoStop}%
\bibitem [{\citenamefont {Samokhin}(2015)}]{Samokhin2015}%
  \BibitemOpen
  \bibfield  {author} {\bibinfo {author} {\bibfnamefont {K.~V.}\ \bibnamefont
  {Samokhin}},\ }\href@noop {} {\bibfield  {journal} {\bibinfo  {journal}
  {Phys. Rev. B}\ }\textbf {\bibinfo {volume} {92}},\ \bibinfo {pages} {174517}
  (\bibinfo {year} {2015})}\BibitemShut {NoStop}%
\bibitem [{\citenamefont {Ito}\ \emph {et~al.}(2016)\citenamefont {Ito},
  \citenamefont {Feng}, \citenamefont {Arita}, \citenamefont {Takayama},
  \citenamefont {Liu}, \citenamefont {Someya}, \citenamefont {Chen},
  \citenamefont {Iimori}, \citenamefont {Namatame}, \citenamefont {Taniguchi},
  \citenamefont {Cheng}, \citenamefont {Tang}, \citenamefont {Komori},
  \citenamefont {Kobayashi}, \citenamefont {Chiang},\ and\ \citenamefont
  {Matsuda}}]{Matsuda2016}%
  \BibitemOpen
  \bibfield  {author} {\bibinfo {author} {\bibfnamefont {S.}~\bibnamefont
  {Ito}}, \bibinfo {author} {\bibfnamefont {B.}~\bibnamefont {Feng}}, \bibinfo
  {author} {\bibfnamefont {M.}~\bibnamefont {Arita}}, \bibinfo {author}
  {\bibfnamefont {A.}~\bibnamefont {Takayama}}, \bibinfo {author}
  {\bibfnamefont {R.-Y.}\ \bibnamefont {Liu}}, \bibinfo {author} {\bibfnamefont
  {T.}~\bibnamefont {Someya}}, \bibinfo {author} {\bibfnamefont {W.-C.}\
  \bibnamefont {Chen}}, \bibinfo {author} {\bibfnamefont {T.}~\bibnamefont
  {Iimori}}, \bibinfo {author} {\bibfnamefont {H.}~\bibnamefont {Namatame}},
  \bibinfo {author} {\bibfnamefont {M.}~\bibnamefont {Taniguchi}}, \bibinfo
  {author} {\bibfnamefont {C.-M.}\ \bibnamefont {Cheng}}, \bibinfo {author}
  {\bibfnamefont {S.-J.}\ \bibnamefont {Tang}}, \bibinfo {author}
  {\bibfnamefont {F.}~\bibnamefont {Komori}}, \bibinfo {author} {\bibfnamefont
  {K.}~\bibnamefont {Kobayashi}}, \bibinfo {author} {\bibfnamefont {T.-C.}\
  \bibnamefont {Chiang}}, \ and\ \bibinfo {author} {\bibfnamefont
  {I.}~\bibnamefont {Matsuda}},\ }\href@noop {} {\bibfield  {journal} {\bibinfo
   {journal} {Phys. Rev. Lett.}\ }\textbf {\bibinfo {volume} {117}},\ \bibinfo
  {pages} {236402} (\bibinfo {year} {2016})}\BibitemShut {NoStop}%
\bibitem [{\citenamefont {Liu}\ \emph {et~al.}(2011)\citenamefont {Liu},
  \citenamefont {Liu}, \citenamefont {Wu}, \citenamefont {Duan}, \citenamefont
  {Liu},\ and\ \citenamefont {Wu}}]{Wu2011}%
  \BibitemOpen
  \bibfield  {author} {\bibinfo {author} {\bibfnamefont {Z.}~\bibnamefont
  {Liu}}, \bibinfo {author} {\bibfnamefont {C.-X.}\ \bibnamefont {Liu}},
  \bibinfo {author} {\bibfnamefont {Y.-S.}\ \bibnamefont {Wu}}, \bibinfo
  {author} {\bibfnamefont {W.-H.}\ \bibnamefont {Duan}}, \bibinfo {author}
  {\bibfnamefont {F.}~\bibnamefont {Liu}}, \ and\ \bibinfo {author}
  {\bibfnamefont {J.}~\bibnamefont {Wu}},\ }\href@noop {} {\bibfield  {journal}
  {\bibinfo  {journal} {Phys. Rev. Lett.}\ }\textbf {\bibinfo {volume} {107}},\
  \bibinfo {pages} {136805} (\bibinfo {year} {2011})}\BibitemShut {NoStop}%
\bibitem [{\citenamefont {Herrmannsdorfer}\ \emph {et~al.}(2011)\citenamefont
  {Herrmannsdorfer}, \citenamefont {Skrotzki}, \citenamefont {Wosnitza},
  \citenamefont {Kohler}, \citenamefont {Boldt},\ and\ \citenamefont
  {Ruck}}]{Ruck2011}%
  \BibitemOpen
  \bibfield  {author} {\bibinfo {author} {\bibfnamefont {T.}~\bibnamefont
  {Herrmannsdorfer}}, \bibinfo {author} {\bibfnamefont {R.}~\bibnamefont
  {Skrotzki}}, \bibinfo {author} {\bibfnamefont {J.}~\bibnamefont {Wosnitza}},
  \bibinfo {author} {\bibfnamefont {D.}~\bibnamefont {Kohler}}, \bibinfo
  {author} {\bibfnamefont {R.}~\bibnamefont {Boldt}}, \ and\ \bibinfo {author}
  {\bibfnamefont {M.}~\bibnamefont {Ruck}},\ }\href@noop {} {\bibfield
  {journal} {\bibinfo  {journal} {Phys. Rev. B}\ }\textbf {\bibinfo {volume}
  {83}},\ \bibinfo {pages} {140501(R)} (\bibinfo {year} {2011})}\BibitemShut
  {NoStop}%
\bibitem [{\citenamefont {Lu}\ \emph {et~al.}(2010)\citenamefont {Lu},
  \citenamefont {Park}, \citenamefont {Yuan}, \citenamefont {Chen},
  \citenamefont {Luo}, \citenamefont {Wang}, \citenamefont {Sefat},
  \citenamefont {McGuire}, \citenamefont {Jin}, \citenamefont {Sales},
  \citenamefont {Mandrus}, \citenamefont {Gillett}, \citenamefont {Sebastian},\
  and\ \citenamefont {Greene}}]{Greene2010}%
  \BibitemOpen
  \bibfield  {author} {\bibinfo {author} {\bibfnamefont {X.}~\bibnamefont
  {Lu}}, \bibinfo {author} {\bibfnamefont {W.~K.}\ \bibnamefont {Park}},
  \bibinfo {author} {\bibfnamefont {H.~Q.}\ \bibnamefont {Yuan}}, \bibinfo
  {author} {\bibfnamefont {G.~F.}\ \bibnamefont {Chen}}, \bibinfo {author}
  {\bibfnamefont {G.~L.}\ \bibnamefont {Luo}}, \bibinfo {author} {\bibfnamefont
  {N.~L.}\ \bibnamefont {Wang}}, \bibinfo {author} {\bibfnamefont {A.~S.}\
  \bibnamefont {Sefat}}, \bibinfo {author} {\bibfnamefont {M.~A.}\ \bibnamefont
  {McGuire}}, \bibinfo {author} {\bibfnamefont {R.}~\bibnamefont {Jin}},
  \bibinfo {author} {\bibfnamefont {B.~C.}\ \bibnamefont {Sales}}, \bibinfo
  {author} {\bibfnamefont {D.}~\bibnamefont {Mandrus}}, \bibinfo {author}
  {\bibfnamefont {J.}~\bibnamefont {Gillett}}, \bibinfo {author} {\bibfnamefont
  {S.~E.}\ \bibnamefont {Sebastian}}, \ and\ \bibinfo {author} {\bibfnamefont
  {L.~H.}\ \bibnamefont {Greene}},\ }\href@noop {} {\bibfield  {journal}
  {\bibinfo  {journal} {Supercond. Sci. Technol.}\ }\textbf {\bibinfo {volume}
  {23}},\ \bibinfo {pages} {054009} (\bibinfo {year} {2010})}\BibitemShut
  {NoStop}%
\bibitem [{\citenamefont {Parab}\ \emph {et~al.}(2017)\citenamefont {Parab},
  \citenamefont {Chauhan}, \citenamefont {Muthurajan},\ and\ \citenamefont
  {Bose}}]{Bose2017}%
  \BibitemOpen
  \bibfield  {author} {\bibinfo {author} {\bibfnamefont {P.}~\bibnamefont
  {Parab}}, \bibinfo {author} {\bibfnamefont {P.}~\bibnamefont {Chauhan}},
  \bibinfo {author} {\bibfnamefont {H.}~\bibnamefont {Muthurajan}}, \ and\
  \bibinfo {author} {\bibfnamefont {S.}~\bibnamefont {Bose}},\ }\href@noop {}
  {\bibfield  {journal} {\bibinfo  {journal} {J. Phys.: Condens. Matter}\
  }\textbf {\bibinfo {volume} {29}},\ \bibinfo {pages} {135901} (\bibinfo
  {year} {2017})}\BibitemShut {NoStop}%
\bibitem [{\citenamefont {Alicea}(2012)}]{Jason2012}%
  \BibitemOpen
  \bibfield  {author} {\bibinfo {author} {\bibfnamefont {J.}~\bibnamefont
  {Alicea}},\ }\href@noop {} {\bibfield  {journal} {\bibinfo  {journal} {Rep.
  Prog. Phys.}\ }\textbf {\bibinfo {volume} {75}},\ \bibinfo {pages} {076501}
  (\bibinfo {year} {2012})}\BibitemShut {NoStop}%
\end{thebibliography}%

\end{document}